\documentclass[final,3p,number,sort&compress]{elsarticle}

\pdfoutput=1

\usepackage[english]{babel}
\usepackage[utf8]{inputenc}
\usepackage[T1]{fontenc}
\usepackage{charter}
\usepackage{setspace}

\usepackage{graphicx,calc}
\usepackage[export]{adjustbox}
\usepackage{color,xcolor}
\definecolor{grau}{HTML}{6F6F6F}

\usepackage{tikz}
\tikzstyle{line}=[draw, thick, -stealth]

\usepackage[font=footnotesize]{caption}
\usepackage[font=footnotesize]{subcaption}
\usepackage{float}
\usepackage{listliketab}

\usepackage{tabularx}
\usepackage{array}
\usepackage{multirow}
\usepackage{booktabs}

\usepackage{amsmath}
\usepackage{amssymb}
\usepackage{amsfonts}
\usepackage{amsxtra}
\usepackage{mathtools}
\usepackage{empheq}
\usepackage{stmaryrd}
\usepackage{icomma}
\usepackage{relsize}

\usepackage[colorlinks=false,hidelinks]{hyperref}

\renewcommand\fbox{\fcolorbox{white!0}{white!0}} 
\newcolumntype{Y}{>{\raggedright\arraybackslash}X}
\newcolumntype{L}[1]{>{\raggedright\let\newline\\\arraybackslash\hspace{0pt}}p{#1}}
\newcolumntype{P}[1]{>{\centering\arraybackslash}m{#1}}

\usepackage{amsfonts}
\usepackage{amssymb}
\usepackage{amsbsy}
\usepackage{ifthen}
\usepackage{multirow}
\usepackage{rotating}
\usepackage{color}
\usepackage{todonotes}
\usepackage{psfrag}
\usepackage{subcaption}
\usepackage{graphicx}

\usepackage{graphicx}
\usepackage[parfill]{parskip}
\usepackage{float}
\usepackage{amssymb, amsmath}
\usepackage{color, fancyhdr}
\usepackage{latexsym, textcomp}
\usepackage{amssymb, amsmath}
\usepackage{lscape}
\usepackage{ae}
\usepackage[T1]{fontenc}
\usepackage{tabularx}
\usepackage[mathscr]{eucal}
\usepackage{longtable}
\usepackage{pstricks}
\usepackage{pst-grad,pst-3d}
\usepackage{color}
\usepackage{enumitem}
\usepackage{epsf,exscale,psfrag,graphics}
\usepackage{xfrac}
\usepackage{placeins}
\usepackage{multirow}

\newcommand{\beq}{\begin{equation}}

\newcommand{\eeq}[1]{%
  \ifthenelse{\equal{#1}{ }}
  {
    \end{equation}
  }
  {
    \label{eq:#1}
    \end{equation}
  }
}

\newcommand{\beqn}      {\begin{eqnarray}}
\newcommand{\eeqn}      {\end{eqnarray}}
\newcommand     {\bea}          {\begin{equationarray}}
\newcommand     {\eea}          {\end{equationarray}}
\newcommand{\beqa}{\begin{eqnarray}}
\newcommand{\eeqa}{\end{eqnarray}}

\newcommand{\bit}       {\begin{itemize}}
\newcommand{\eit}       {\end{itemize}}

\newcommand{\bde}       {\begin{description}}
\newcommand{\ede}       {\end{description}}

\newcommand{\bec}       {\begin{center}}
\newcommand{\eec}       {\end{center}}

\newsavebox{\BildText}
                        {\centerline{\usebox{\BildText}}\end{figure}}
                        
\newcommand{\Kasten}[3]%
{\begin{equation}
  \fbox{
    \begin{minipage}{5.5in}
      \centering
      #2
      \rule{5.4in}{0.5pt}
      {\bf #3}
    \end{minipage}
  }
  \label{box:#1}
\end{equation}}

\newcommand{\RefBraces}[1]%
{%
  \typeout{Referenz #1 in Seite \thepage}%
  \ifthenelse{\equal{\pageref{#1}}{\thepage}}%
             {(\ref{#1})}%
             {(\ref{#1}) auf Seite \pageref{#1}}%
}

\DeclareMathAlphabet\eufm{U}{euf}{m}{n}
\DeclareMathAlphabet\eufb{U}{euf}{b}{n}
\DeclareMathAlphabet\eurm{U}{eur}{m}{n}
\DeclareMathAlphabet\eurb{U}{eur}{b}{n}
\DeclareMathAlphabet\eusm{U}{eus}{m}{n}
\DeclareMathAlphabet\eusb{U}{eus}{b}{n}
\DeclareMathAlphabet\mathsfm{OT1}{cmss}{m}{n}
\DeclareMathAlphabet\mathsfb{OT1}{cmss}{bx}{n}

\newcommand{\Qed}{\hfill \mbox{$\Box$}}         

\newcommand{\bigo}[1]{\setbox0\hbox{$\bigcirc$}
             \rlap{\raise .2ex\hbox to \wd0{\hfil ${\scriptscriptstyle
                   #1}$\hfil}}\box0}
 
\newcommand{\subf}[2]{%
	{\small\begin{tabular}[t]{l@{}c@{}}
			#1\\#2
	\end{tabular}}%
}
\newcommand{\svektor}[2]%
{\left(\!\!\!\begin{array}{c} #1\\#2 \end{array}\!\!\!\right)}

\newcommand{\zvektor}[2]%
{\left(#1,#2\right)^T}

\newcommand{\Svektor}[2]%
{\left[\!\!\!\begin{array}{l} #1\\#2 \end{array}\!\!\!\right]}

\newcommand{\Zvektor}[2]%
{\left[#1,#2\right]^T}

\newcommand{\Rho}{R}

\journal{arXiv}

\begin{document}
	
\begin{frontmatter}
		
\title{Modelling effects of moisture on mechanical properties of crosslinked polyurethane adhesives}

\author[add1]{S. P. Josyula 
}\ead{siva.josyula@uni-saarland.de}
\author[add2]{ M. Brede  }
\author[add2]{O. Hesebeck }
\author[add2]{ K. Koschek  }
\author[add3]{ W. Possart }
\author[add2]{ A. Wulf  }
\author[add3]{   B. Zimmer  }
\author[add1]{S. Diebels 
}
\ead{s.diebels@mx.uni-saarland.de}
		
\address[add1]{Chair of Applied Mechanics, Saarland University, Saarbr\"ucken, Germany}
\address[add2]{Fraunhofer Institute for Manufacturing Technology and Advanced Materials, Wiener Stra{\ss}e 12, Bremen, Germany}
\address[add3]{Chair for Adhesion and Interphases in Polymers, Saarland University, Saarbr\"ucken, Germany}
	\begin{abstract}
			Crosslinked polyurethane adhesives show large deformation viscoelastic behaviour and age under the moisture influence because of their hygroscopic behaviour. The viscoelastic behaviour of the material is modelled with the micromechanical network model. The micromechanical model considers the shorter and longer chains with a probability distribution function. The network evolution concept is used to model softening of material due to the breakage of the shorter chains with an increase in deformation. The moisture diffusion in the polyurethane adhesive is  behaviour, therefore Langmuir-type diffusion model is used to model moisture diffusion. The transported moisture in the material leads to an exponential decay in the mechanical properties causing the ageing of the material. The micromechanical model needs to be coupled with the Langmuir-type diffusion model to analyse the ageing process, where the mechanical properties are considered as the function of the local moisture concentration.
	\end{abstract}
\end{frontmatter}
\section{Introduction}
Due to the relatively favourable stress state in glue joints, the glueing technology is predestined for joining thin-walled lightweight structures in vehicle and plant construction. However, the safety-relevant structural bonding of primary structures is generally avoided, as it is not yet possible to calculate the service life of the glue joints under the influence of water or other environmental media. This means that resource-saving lightweight construction potentials are being given away to a large extent. With the increasing use of fiber composite plastics in automotive manufacturing, higher-strength, hyperelastic polyurethane (PU) adhesives with a glass transition temperature of approx. -20°C and a wide glass transition area (up to 60K) are available, which are very well suited for structural connections of this type. In a similar way, PU adhesives with glass transition are used in the area around 30°C. Understanding the long-term durability of adhesive bonds is a significant research focus because of their viscoelastic behaviour and their sensitivity to the surrounding atmospheric conditions. Therefore, it is necessary to study the effects of environmental conditions on the mechanical behaviour of the adhesive. Recently efforts have been made to study mechanical behaviour under the influence of temperature, and moisture diffusion \cite{MEI08,HUA08,MUB09,BRE15}. In the research, the micromechanical behaviour is not considered in the numerical simulation. 

Material models to evaluate viscoelastic behaviour are classified into phenomenological and micromechanical-based network models. The phenomenological model is used to describe the macro mechanical behaviour of the material. These models are formulated from the empirical relation derived from the experimental data and observation but not supported by theory. The parameters of phenomenological models are identified by fitting the experimental data and have no relevance to the molecular structure of the material. Mooney-Rivlin \cite{Mooney,Rivlin}, Yeoh \cite{Yeoh}, and Ogden \cite{Ogden} models are some of the popular phenomenological models based on an invariant or principal stretch of the macroscopic continuum theories. On the contrary, the micromechanical network models are formulated based on the statistical chain mechanics with a motivation to describe the complex micromechanical behaviour. 3-chain model \cite{JAM43,3chain1}, the eight-chain model of Arruda and Boyce \cite{ARR93} are the popular material models implemented in commercial finite element programs. A non-affine network model that includes the orientation of chains in the sphere was presented by Miehe et al. \cite{MIE04}. These models idealise the chain distribution of equal lengths and do not consider softening behaviour of the material. 

The microstructure of a crosslinked polymer network is formed of shorter and longer chain distribution, and the shorter chains break/debond at a smaller stretch leading to stress softening \cite{Bueche}. Experiments performed on the crosslinked polymers show stress softening is because of breaking/debonding of the chains of the polymer network \cite{Harwood}. Numerical modelling of the stress-softening behaviour of crosslinked polymer materials has been active research for a long time. Where classical Gaussian or non-Gaussian chain statistics are applied to model the elastic and inelastic behaviour of the polymers. The Gaussian chain statistics uses the end-to-end distance of a single chain formed of a fixed number of chain segments by considering the exact distribution of chain \cite{TRE46}. In comparison, non-Gaussian chain statistics considers a freely joined chain expressed with inverse Langevin function \cite{Kuhn1942}. Govindjee et. al. \cite{Govindjee}, Smeulders et. al. \cite{SME99} proposed a model based on Arruda–Boyce network model that accounts for the molecular weight distribution to consider the chain length distribution in the polymer network. Marckmann et al. \cite{Marckmann} proposed a softening network model by altering the Arruda–Boyce eight-chain network model and considering a mean number of chain segments in a polymer network. G\"oktepe et. al. \cite{Goektepe} developed a micromechanical model considering a non-affine approach to include Mullins-type damage due to breakage/debonding of chains in a network. This non-affine micromechanical model is based on the numerical integration scheme proposed by Bažant et.al \cite{Bazant}. In the aforementioned micromechanical softening models, the softening due to damage of the polymer chains is considered with a phenomenological damage function based on the history variables. Dargazany et.al. \cite{Dargazany} extended the Govindjee et. al. \cite{Govindjee} model using a numerical integration scheme \cite{Bazant} to include anisotropic behaviour in carbon-filled rubber and softening due to damage in chains is governed by the network evolution. Recently Itskov et al. \cite{ITS16} proposed a full network rubber elasticity and softening model based on the numerical integration over a unit sphere discussed earlier \cite{Bazant}. This model does not consider the filler particles, and the softening behaviour is motivated by the network evolution with an assumption that the distribution of chain segments increases with the maximum stretch.

The present work investigates the finite-strain viscoelastic behaviour of the crosslinked polyurethane adhesive with a softening-based micromechanical model. The micromechanical model is used here to consider the softening of the material with an increase in the stretch. This model considers shorter and longer chain length distribution in a random network of polymer chains. The damage in the chains with an increase in the stretch is modelled based on the network evolution theory. The model discussed here is isotropic and does not include deformation-induced anisotropy. As discussed earlier, the crosslinked polyurethane adhesives are hygroscopic and absorb moisture from the atmosphere causing the material properties to decay, thus leading to early ageing. The ageing in adhesives due to moisture transport is reversible, therefore the chemical ageing of material is not treated in the numerical modelling. Experimental investigation \cite{HUA16} of moisture transport in the crosslinked polyurethane adhesive is anomalous, leading to diffusion of moisture characterised into mobile and immobile moisture concentrations. The anomalous diffusion of moisture is modelled with Langmuir-type diffusion \cite{CAR78}. Numerical investigation of moisture diffusion with the Langmuir-type diffusion model has already been investigated in epoxy-based adhesive \cite{POP05,AME10}, and these investigations efficiently explain the presence of mobile and immobile moisture concentration in the material domain.

It is necessary to couple the mechanical behaviour with the diffusion behaviour to model the ageing behaviour under the moisture influence. Roy et. al. \cite{Roy} investigated the influence of moisture or solvent diffusion on the non-linear viscoelastic behaviour, where the non-linear Fickian behaviour is taken into consideration, and temperature-dependent diffusion coefficient is used. The influence of diffusion behaviour is investigated on incompressible polymer gels to understand the local rearrangement of molecules due to swelling of gels under large deformations by Hong et. al. \cite{Hong}. Many more coupled material models \cite{Chester,Haghighi,Su} has been developed to investigate the effects of diffusion on the deformation in the polymer gels within the framework of the finite-strain theory. Recently, Goldschmidt et. al. \cite{Goldschmidt} has numerically investigated the mechanical behaviour of crosslinked polyurethane under the influence of the humid atmosphere. In which the finite-strain viscoelastic model is coupled with the Fick diffusion using an exponential decay function. Sharma et. al. \cite{Sharma} numerically investigated the moisture transport behaviour in polyamide by considering moisture-dependent material properties, and moisture diffusion is modelled using Fick's law. None of the aforementioned theories considers micromechanical and anomalous moisture diffusion behaviour. In the present paper, the proposed micromechanical viscoelastic model is coupled with the Langmuir-type diffusion model and moisture-dependent material parameters of the mechanical model are considered to model ageing behaviour analogous to Sharma et. al. \cite{Sharma}.
\section{Micromechanically motivated polymer free energy}\label{micromech}
The basic idea to model a crosslinked polymer network material is to consider shorter and longer chain distribution in a random network, and the shorter chains tend to break/debond from the network leading to softening of the material. This idea is accounted in the micromechanical material model formulation by multiplying isotropic free energy ${\rm \Psi}$ of the material with a cumulative distribution function $G(\lambda_m)$ 
\begin{equation}
	W({\rm{I}_1},\lambda_c)={\rm \Psi}\int\limits_{1}^{\infty}g(\lambda_m){\rm{d}}\lambda_m = {\rm \Psi}G(\lambda_m).
	\label{eq:free2-0}
\end{equation}
The cumulative distribution function considers the break/debonding of the shorter chains following the evolution network theory. In this theory, the shorter chains becomes inactive with increase in the deformation and the processes is assumed to be irreversible to consider the softening of the material. The cumulative distribution function is derived from the probability distribution function defined for a random network. The probability distribution function is defined as a function of the current chain stretch of an ideal chain to evaluate the statistical information concerning the stretch in a random network. 
\subsection{Polymer chain terminology}
An individual chain is formed from $N$ segments of chain links of a uniform length $l\,\rm mm$. The end-to-end distance $r_0$ between endpoints of an ideal chain is arbitrarily calculated as \cite{James,Kuhn1942,Kuhn1936}  
\begin{equation}
	r_0=\sqrt{N}l,
\end{equation}
and the maximum length of chain $r_m$ is calculated as
\begin{equation}
	r_m = Nl.
\end{equation}
The current chain stretch under strained conditions is calculated as 
\begin{equation}
	r_c=\frac{1}{\sqrt{3}}\sqrt{N}l\sqrt{\rm{I}_1},
	\label{eq:lcs1}
\end{equation}
where ${\rm{I}_1}$ is the first invariant of left Cauchy-Green deformation tensor. Current chain stretch $\lambda_c$, and maximum chain stretch $\lambda_{\rm max}$ are calculated as
\begin{equation}
	\lambda_c=\frac{r_c}{r_0}=\frac{\sqrt{\rm{I}_1}}{\sqrt{3}}, \hspace{2 mm}\text{and} \hspace{2 mm} \lambda_{\rm max} = \frac{r_m}{r_0} = \sqrt{N}.
	\label{eq:lcs4}
\end{equation}
The current chain stretch is defined in the interval between initial chain stretch $\lambda_0$ of a rigid chain and maximum $\lambda_{\rm{max}}$ chain stretch 
\begin{equation}
	\lambda_0=1\le \lambda_c = \frac{\sqrt{\rm{I}_1}}{\sqrt{3}} \le \lambda_{\rm{max}}.
	\label{eq:lcs6}
\end{equation}
\subsection{Probability distribution function}
\begin{figure}[H]
	\centering
	\def\svgwidth{.7\textwidth}
	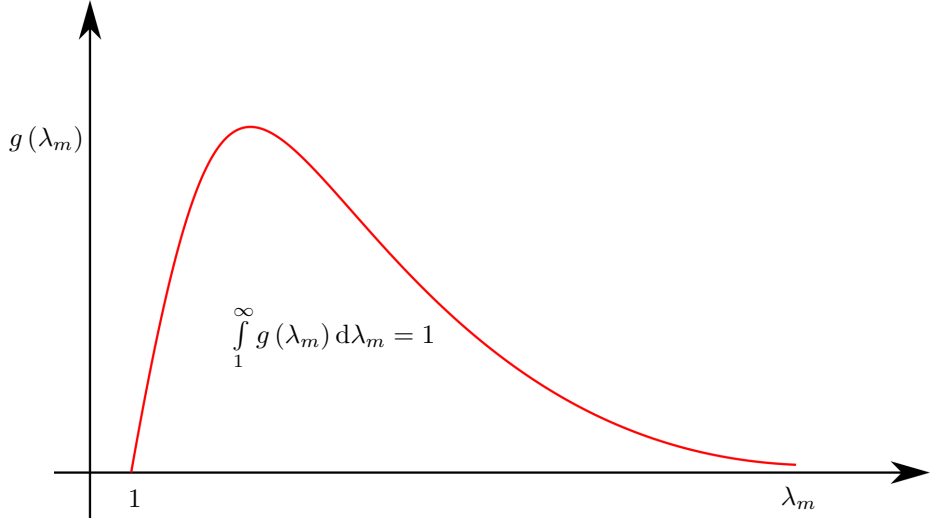
	\caption{Shorter and longer chain stretch distribution for an unstrained material}	
	\label{unstr}
\end{figure}

\begin{figure}[H]
	\centering
	\def\svgwidth{0.7\textwidth}
	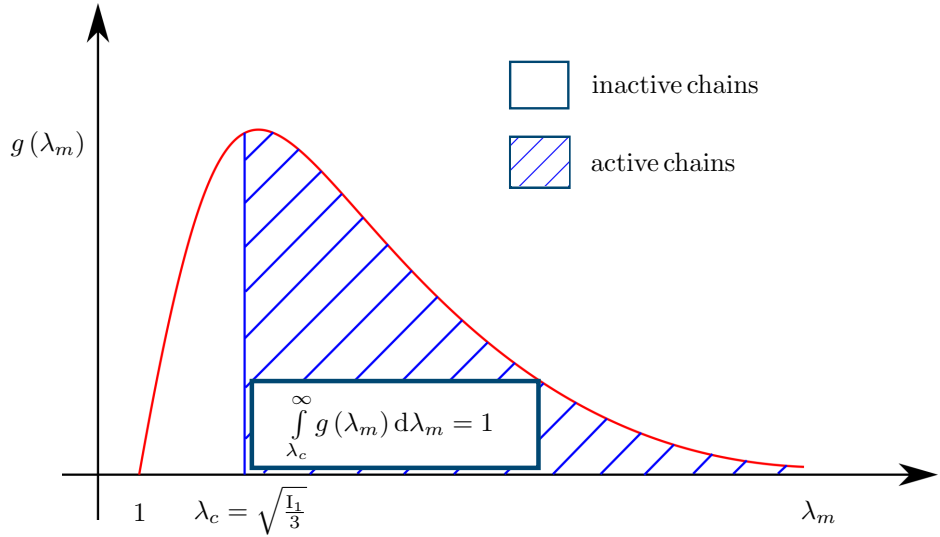
	\caption{Distribution of active and inactive chain of a deformed material}	
	\label{dampol}
\end{figure}

The maximum chain stretch distribution is defined as analogous to the Wesslau mass distribution function by replacing the mass variable with the actual chain stretch. The Probability distribution function $g(\lambda_m)$ is evaluated from the maximum chain stretch distribution $\rm{w}(\lambda_m)$ as 
\begin{equation}
	\begin{aligned}
		g(\lambda_m) = \frac{^N\lambda_m}{(\lambda_m-1)}{\rm{w}(\lambda_m)}= \frac{^N\lambda_m}{\beta \sqrt{\pi}}\frac{1}{(\lambda_m - 1)^2} {\rm{exp}} \left(\frac{-1}{\beta^2}\left({\rm{ln}}\left(\frac{(\lambda_m -1 )}{^0 \lambda_m}\right)\right)^2\right).
	\end{aligned}
	\label{eq:lcs18}
\end{equation}
For simplicity, $g(\lambda_m)$ is rearranged as follows 
\begin{equation}
	g(\lambda_m) = a_0\frac{1}{(\lambda_m-1)^2} \rm{exp}\left(-a_1 \left( \rm{ln}\left(a_2(\lambda_m-1) \right) \right)^2 \right)
	\label{eq:lcs20}
\end{equation}
with parameters 
\begin{equation}
	a_0 = \frac{^N\lambda_m}{(\beta\sqrt{\pi})};\, a_1=\frac{1}{\beta^2}\,{\rm and}\, a_2 = \frac{1}{^0\lambda_m}.
	\label{param}
\end{equation}
The constants $\beta$, $^0\lambda_m$ and $^N\lambda_m$ are calculated with the help of the material parameters polydispersity index $Q$ and average chain elongation $^M\lambda_m$
\begin{equation}
	\beta=\sqrt{2{\rm{ln}(Q)}};\, ^N\lambda_m = \frac{^M\lambda_m}{Q};\, ^0\lambda_m = \sqrt{^M\lambda_m \, ^N\lambda_m}. 
\end{equation}
In an unstrained material, every chain actively participates in the free energy of the material. Therefore, the function $g(\lambda_m)$ is integrated between the interval $[1, \,\infty]$ to compute cumulative distribution function 
\begin{equation}
	G\left(\lambda_m\right) = \int\limits_{1}^{\infty}g\left(\lambda_m\right) = 1,
\end{equation}
where $\infty$ is the maximum chain stretch $\lambda_m$. The schematic representation of the function $g(\lambda_m)$ for an unstrained material is shown in Fig. \ref{unstr}.
Softening due to breaking/debonding of chains in a random network starts with the shorter chains eventually leading to breaking the long chains with a increase in the load. The breakage in the smaller chains under applied load is explained with a schematic diagram shown in Fig. \ref{dampol}. Here, a random network is considered to experience a current stretch of $\lambda_c$ leading to the breaking of shorter chains represented as inactive chains. The other chains that contribute to the mechanical free energy are active chains.
The cumulative distribution function of the active chains under a current stretch $\lambda_c$ follows the integration of $g\left(\lambda_m\right)$ at an interval $\left[\lambda_c \, \infty\right]$ 
\begin{equation}
	G\left(\lambda_m\right) = \int\limits_{\lambda_c}^{\infty}g\left(\lambda_m\right),
	\label{cumu1}
\end{equation}
where the maximum chain elongation ($\infty$) of a material is a priori unknown quantity, therefore the cumulative density function of the active chains given in the equation \eqref{cumu1} is reformulated by subtracting the cumulative distribution function of the inactive chains from the cumulative distribution function of the unstrained material. As a result, the cumulative distribution function of the active chains follows 
\begin{equation}
	\begin{aligned}
		G\left(\lambda_m\right) =\int\limits_{1}^{\infty}g(\lambda_m){\rm{d}}\lambda_m - \int\limits_{1}^{\lambda_c} g(\lambda_m) {\rm{d}}\lambda_m = 1 - \int\limits_{1}^{\lambda_c} g(\lambda_m) {\rm{d}}\lambda_m.
		\label{eq:free2}
	\end{aligned}
\end{equation}
The indefinite integral of the probability density function is evaluated to obtain the cumulative distribution function. After some mathematical evaluation, the cumulative distribution function dependent on $\lambda_m$ is derived as the function of an error function to avoid non-elementary integrals.
\begin{equation}
	G(\lambda_m) = \frac{a_0 a_2 {\rm{exp}}\left(\frac{1}{4a_1} \right)\sqrt{\pi}}{2\sqrt{a_1}} {\rm{erf}}\left( \frac{1+2a_1  \rm{ln}\left(a_2(\lambda_m-1) \right)}{2\sqrt{\pi}}\right) + C.
	\label{eq:lcs21}
\end{equation}
The integration constant $C$ is determined with an assumption $G(\lambda_m = 1) =0$ and the error function value as $-1$ to avoid singularity due to logarithmic term. As a result, the integration constant is derived as 
\begin{equation}
	C = \frac{a_0 a_2 \rm{exp}\left(\frac{1}{4a_1} \right)\sqrt{\pi}}{2\sqrt{a_1}},
	\label{eq:lcs22}
\end{equation}
by inserting equation \eqref{eq:lcs22} in \eqref{eq:lcs21} leads to the cumulative density distribution function as:
\begin{equation}
	G(\lambda_m) = \frac{a_0 a_2 \rm{exp}\left(\frac{1}{4a_1} \right)\sqrt{\pi}}{2\sqrt{a_1}}\left(1+ \rm{erf}\left( \frac{1+2a_1  \rm{ln}\left(a_2(\lambda_m-1) \right)}{2\sqrt{\pi}}\right)\right).
	\label{eq:lcs23}
\end{equation}
After inserting equation \eqref{eq:free2} in \eqref{eq:free2-0}, the micromechanical free energy density function is derived for a current stretch $\lambda_c$ in a random network as
\begin{equation}
	\begin{aligned}
		W({\rm{I}}_1,\lambda_c) ={\rm \Psi}({\rm{I}}_1)\left(\int\limits_{1}^{\infty}g(\lambda_m){\rm{d}}\lambda_m - \int\limits_{1}^{\lambda_c({\rm{I}}_1)} g(\lambda_m) {\rm{d}}\lambda_m \right) ={\rm \Psi}({\rm{I}}_1)\left(1 - \int\limits_{1}^{\lambda_c({\rm{I}}_1)} g(\lambda_m) {\rm{d}}\lambda_m \right),
	\end{aligned}
	\label{energy}
\end{equation}
where Neo-Hooke model is considered as a free energy ${\rm \Psi}({\rm{I}}_1)$ in softening micromechanical model for simplicity. However, the crosslinked polyurethane adhesive shows a nearly incompressible viscoelastic behaviour. Therefore, the energy function \eqref{energy} is uniquely decoupled into volume and shape-changing parts based on continuum mechanical description \cite{Marckmann,Miehe1994} to further extend the defined material model to analyse finite-strain viscoelastic behaviour.
\section{Coupled material model}\label{viscoelas}
The micromechanical network model is extended to viscoelastic behaviour using a rheological description consisting of Maxwell elements connected in parallel to the spring element. The deformation in the body is calculated based on the continuum mechanics of finite-strain theory \cite{HauptL}.
\begin{figure}[H]
	\centering
	\vspace{4mm}
	\def\svgwidth{0.8\textwidth}
	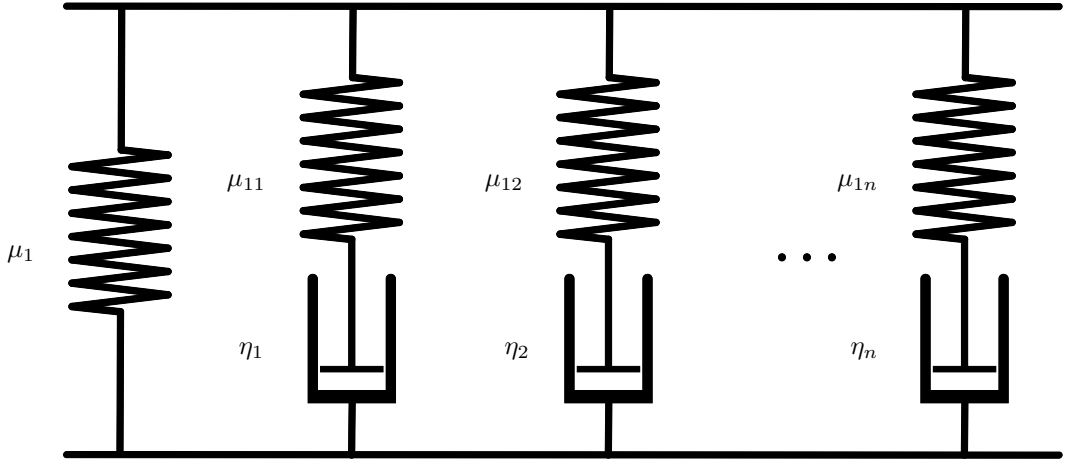
	\vspace*{2mm}	
	\caption{Rheological model of the viscoelasticity with $n$ Maxwell elements.}
	\label{finviskoelast40}
\end{figure}
\subsection{Kinetics of finite-strain theory}
The deformation gradient tensor $\mathbf{F}$ needs to be decomposed into elastic $\mathbf{F}_e^j$ and inelastic $\mathbf{F}_i^j$ deformation gradient tensor of $j=1,...,n$ Maxwell element \cite{Lee} to model the viscoelastic material model. The deformation tensor is decomposed with multiplicative decomposition
\begin{equation}
	\mathbf{F}\,=\,\mathbf{F}_e^j\cdot\mathbf{F}_i^j\,.
	\label{finvisko390}
\end{equation}
The free energy is decomposed into shape and volume-changing parts to model a nearly incompressible viscoelastic behaviour. This additive decomposition of free energy leads to multiplicative decomposition of the deformation gradient tensor $\mathbf{F}$ into shape and volume-changing parts and is given as
\begin{equation}
	\mathbf{F}=\mathbf{F}_{\rm vol}\cdot\mathbf{F}_{\rm iso},
	\label{decom}
\end{equation}
where $\mathbf{F}_{\rm vol}$ as volumetric part and $\mathbf{F}_{\rm iso}$ as the isochoric part of the deformation gradient tensor. The volumetric $\mathbf{F}_{\rm vol}$ and isochoric $\mathbf{F}_{\rm iso}$ parts of the deformation tensor is calculated as
\begin{equation}
	\mathbf{F}_{\rm iso} = J^{-1/3} \mathbf{F}, \hspace{1cm}\mathbf{F}_{\rm vol} = J^{1/3}\mathbf{I}
	\label{isodefo}
\end{equation}
with the Jacobian $J={\rm det}\mathbf{F}$ and second-order identity tensor $\mathbf{I}$. The right Cauchy-Green deformation tensor $\mathbf{C}=\mathbf{F}^T\cdot\mathbf{F}$ is reformulated to isochoric right Cauchy-Green deformation tensor $\bar{\mathbf{C}}$ as $\bar{\mathbf{C}}=J^{-2/3}\mathbf{C}$ from the isochoric deformation gradient tensor given in equation \eqref{isodefo} \cite{Miehe1994}. The first ${\rm I}_1$ and third ${\rm I}_3$ invariants of the Cauchy-Green deformation tensor are calculated as
\begin{equation}
	{\rm I}_1 = {\rm tr}\left(\mathbf{C}\right) = {\rm tr}\left(\mathbf{B}\right);\,\,
	{\rm I}_3 = {\rm det}\left(\mathbf{C}\right) = {\rm det}\left(\mathbf{B}\right) = J^2,
\end{equation}
where $\mathbf{B}=\mathbf{F}\cdot\mathbf{F}^T$ is the left Cauchy-Green deformation tensor. The counterparts of the invariants are calculated as
\begin{equation}
	\bar{{\rm I}}_1 = J^{-2/3} {\rm I}_1 \hspace{1cm} {\rm and} \hspace{1cm} \bar{{\rm I}}_3=1
\end{equation}
By combining equations \eqref{finvisko390}, \eqref{decom} and \eqref{isodefo}, the kinetics of the elastic part of the Maxwell element is formulated as
\begin{equation}
	\mathbf{\bar{B}}_{e}^j = (\mathbf{F}_e^j)^{(\rm iso)} \cdot ((\mathbf{F}_e^j)^{(\rm iso)})^T =\mathbf{F}_{\rm iso}\cdot(\mathbf{\bar{C}}_{i}^j)^{-1} \cdot \mathbf{F}_{\rm iso}^T\,.
	\label{finvisko400}
\end{equation}
\subsection{Thermodynamic consistency}
The mechanical behaviour of polyurethane adhesive is dependent on the material's moisture transport. Therefore, the free energy is additively decomposed into the mechanical and diffusion parts, $W_{\rm mech}$ and $W_m$ 
\begin{equation}
	W = W_{\rm mech}\left(J, {\rm I}_1^{\bar{\mathbf{B}}},{\rm I}_1^{{\bar{\mathbf{B}}}^j_e}, \lambda_m ,m\right) \hspace{1mm}+ W_m(m,m_b).
	\label{fullener}
\end{equation}
Polyurethane adhesive is considered a nearly incompressible material because of the volumetric strains, therefore an uncoupled response is considered to define the free energy function \cite{HARTMANN20021439,SIMO1987153}. The uncoupled mechanical response is based on the additive decomposition of the free energy into the volumetric and isochoric parts. The isochoric part considers the equilibrium part and the rate-dependent non-equilibrium part. The moisture-dependent function is expressed as 
\begin{equation}
	\!W_{\rm mech} = W_{\rm vol} \left(J,\lambda_m,m \right)+ W_{\rm eq} \left(\bar{\rm I}_1^{\bar{\mathbf{B}}}, \lambda_m, m \right) +  \sum_{j=1}^{n} W^j_{\rm neq}\left({\rm I}_1^{\bar{\mathbf{B}}^j_e},\lambda_m, m\right),
	\label{finvisko450-1}
\end{equation}
where $W_{\rm vol}$ is the volumetric free energy and $W_{\rm eq}$ and $W_{\rm neq}^j$ are the equilibrium and the non-equilibrium parts of the isochoric part. The non-equilibrium part is comprised of $j=1,2,...,n$ Maxwell elements. After substituting equation \eqref{finvisko450-1} leads to the coupled free energy
\begin{equation}
	\begin{aligned}
		W = W_{\rm vol}(J,\lambda_m,m) + W_{\rm eq}\left({\rm I}_1^{{\mathbf{\bar{B}}}},\lambda_m,m \right)
		+ \sum_{j=1}^{n} W^j_{\rm neq}\left({\rm I}_1^{\bar{ \mathbf{B}}^j_e},\lambda_m,m \right) \hspace{2mm}+   W_m(m,m_b).
	\end{aligned}
\end{equation}
An important requirement for the material model is to satisfy the Clausius-Planck inequality. The Clausius-Planck inequality for the coupled diffusion and mechanical behaviours at the isothermal condition follows
\begin{equation}
	\rho\dot{W} - \mathbf{T}\colon \mathbf{D} + {\rm div} \left(\Rho_m \mathbf{q}\right) \geq 0,
	\label{clausiu}
\end{equation}
where $\Rho_m$ is the chemical potential and $\mathbf{q}$ is the moisture flux. The process variables for the defined coupled material model are
\begin{equation}
	{\cal S} = \left\{\mathbf{B},\mathbf{B}_e^j, m, {\rm grad}\, m\right\},
\end{equation}
and the constitutive quantities are 
\begin{equation}
	{\cal R} = \left\{W,\mathbf{T},\mathbf{q}\right\}.
\end{equation}
The material time derivative of the free energy function is derived with the process variable to evaluate dissipation as 
\begin{equation}
	\begin{aligned}
		\dot{W} = \frac{\partial W_{\rm vol}(J,\lambda_m,m)}{\partial \bar{ \mathbf{B}}} \colon \dot{ \mathbf{B}} + \frac{\partial W_{\rm eq} \left({\rm I}_1^{{\bar{ \mathbf{B}}}},\lambda_m,m \right)}{\partial \bar{ \mathbf{B}}} \colon \dot{ \mathbf{B}} +\sum_{j=1}^{n} \frac{\partial W^j_{\rm neq}\left({\rm I}_1^{{\bar{ \mathbf{B}}}^j_e},\lambda_m,m \right)}{\partial {\bar{ \mathbf{B}}}^j_e} \colon \dot{{\mathbf{B}}}^j_e + \frac{\partial W_{m}\left(m,m_b \right)}{\partial m} \colon \dot{m}.
	\end{aligned}\label{freeder}
\end{equation}
The time derivatives of the deformation tensors are formulated with the deformation velocity $\mathbf{D}$ as 
\begin{equation}
	\dot{\mathbf{B}} = 2\mathbf{D}\cdot\mathbf{B} \hspace{3mm} {\rm and} \hspace{3mm}
	\dot{\mathbf{B}}_e^j = 2\mathbf{D}\cdot\mathbf{B}_e^j - 2\mathbf{F}_e^j\cdot \hat{\mathbf{\Gamma}}_i^j \cdot \left(\mathbf{F}_e^j\right)^T,
\end{equation} 
where the inelastic deformation rate of the intermediate configuration $\hat{\mathbf{\Gamma}}_i^j$ is an outcome of the product rule applied over $\dot{\mathbf{B}}_e^j$. The Clausius-Planck inequality is derived with the material time derivative of the free energy \eqref{freeder} and the time derivatives of the deformation tensor as 
\begin{equation}
	\begin{aligned}
		&\left(-2\rho \mathbf{B}\cdot \frac{\partial W_{\rm vol}}{\partial \mathbf{B}} -2\rho{\mathbf{B}}\cdot\frac{\partial{W_{\rm eq}}}{\partial{\mathbf{B}}} - \sum_{j=1}^{n}2\rho\bar{\mathbf{B}}_e^j\cdot\frac{\partial{W_{\rm neq}}} {\partial\bar{\mathbf{B}}_e^j} + \mathbf{T}\right)\colon \mathbf{D}\,\\
		+\,
		&\hspace{1mm}\sum_{j=1}^{n}2\rho\frac{\partial{W_{\rm neq}}}{\partial\bar{\mathbf{B}}_e^j} : \left(\mathbf{F}_e^j\cdot \hat{\mathbf{\Gamma}}_i^j \cdot \left(\mathbf{F}_e^j\right)^T \right) +\,\left(-\rho\frac{\partial W_{m}(m,m_b)}{\partial m} + \Rho_m\right)\cdot\dot{m}
		- {\rm grad} \,\Rho_m \cdot\mathbf{q} \geq 0.
	\end{aligned}
	\label{inequality}
\end{equation}
The inequality \eqref{inequality} depends linearly on the independent variables $\mathbf{D}$, $\dot{m}$. Thus leading to the stress and the diffusion potential from Coleman and Noll \cite{Coleman} argumentation
\begin{equation}
	\begin{aligned}
		\mathbf{T} =\, &2\rho \mathbf{B}\cdot \frac{\partial W_{\rm vol}}{\partial \mathbf{B}} +2\rho{\mathbf{B}}\cdot\frac{\partial{W_{\rm eq}}}{\partial{\mathbf{B}}} + \sum_{j=1}^{n}2\rho\bar{\mathbf{B}}_e^j \cdot \frac{\partial{W_{\rm neq}^j}} {\partial\bar{\mathbf{B}}_e^j},\\
		\Rho_m =& \rho \frac{\partial W_{m}(m,m_b)}{\partial m}.
	\end{aligned} \label{constui}
\end{equation}
The first term of the stress tensor represents the volumetric stress analogues to the hydrostatic stress \cite{SIMO1987153}, and the other two terms correspond to the equilibrium and non-equilibrium parts. Applying the chain rule to the volumetric stress component with the relationship $\partial J/\partial\mathbf{B} = J\mathbf{B}^{-1}$ and assuming $W_{\left(\bullet\right)} = \rho W_{\left(\bullet\right)}$ \cite{HARTMANN20021439} leads to the stress tensor
\begin{equation}
	\begin{aligned}
		\mathbf{T} =\, J W_{\rm vol}'\mathbf{I} +2{{\mathbf{B}}}\cdot\frac{\partial{W_{\rm eq}}}{\partial{\mathbf{B}}} + \sum_{j=1}^{n}2\bar{\mathbf{B}}_e^j\cdot\frac{\partial{W_{\rm neq}^j}} {\partial\bar{\mathbf{B}}_e^j},
		\label{totalstress}
	\end{aligned}
\end{equation}
where $W'_{\rm vol} = \partial W_{\rm vol}/\partial J$. The dissipation inequality is simplified to 
\begin{equation}
	\begin{aligned}
		\sum_{j=1}^{n}2\rho\frac{\partial{W_{\rm neq}^j\left({\rm I}_1^{{\mathbf{B}}^j_e},m \right)}}{\partial\bar{\mathbf{B}}_e^j} : \left(\mathbf{F}_e^j\cdot \hat{\mathbf{\Gamma}}_i^j \cdot \left(\mathbf{F}_e^j\right)^T \right) - {\rm grad}\Rho_m\cdot\mathbf{q} \geq 0. 
	\end{aligned}\label{simplediss1}
\end{equation}
Using the kinematic relations and applying tensor relations, the first term of the dissipation inequality of equation \eqref{simplediss1} results in the evolution equation of the right Cauchy-Green deformation \cite{Haupt,Lion,Seldan}
\begin{equation}
	\begin{aligned}
		\dot{\bar{\mathbf{C}}}_i^j\,&=\,\frac{4}{r_j} \left[\bar{\mathbf{C}} - \frac{1}{3}{\rm{tr}}\left(\bar{\mathbf{C}}\cdot\left(\bar{\mathbf{C}}_i^j\right)^{-1} \right)\bar{\mathbf{C}}_i^j\right]\\
	\end{aligned}
	\label{inelasticright}
\end{equation} 
where $r_j$ is the relaxation time associated with $j^{th}$ Maxwell element. The relaxation times are the material constants introduced as 
\begin{equation}
	r_j=\frac{\mu_{10n}}{\eta_n}.
\end{equation}
The second term is characterised by the diffusive flux to ensure the positivity of the simplified dissipation inequality 
\begin{equation}
	\mathbf{q} = - D \left({\rm grad}\, \Rho_m\right),
	\label{flux}
\end{equation}
where $D$ is the diffusion coefficient. Free energy of the moisture diffusion $W_m(m,m_b)$ is defined for the anomalous moisture diffusion as 
\begin{equation}
	W_m(m,m_b) = \frac{1}{2}\left(m-m_b\right)^2,	
	\label{moisfree}
\end{equation}
where $m$ and $m_b$ are the total and immobile moisture concentrations. The chemical potential equation derived in the equation \eqref{constui}, diffusive flux \eqref{flux}, the diffusion free energy equation \eqref{moisfree} and the balance of mass \cite{Haupt,Holzapfel} leads to the anomalous diffusion equation
\begin{equation}
	\frac{{\rm d} m}{{\rm d} t} = D\,{\rm div}\left({\rm grad}\, (m-m_b)\right),
\end{equation}
and the immobile moisture concentration $m_b$ is calculated with an evolution equation \cite{CAR78}
\begin{equation}
	\dot{m_b}= \alpha m_f - \beta m_b.
	\label{eq:diff2}
\end{equation}
The symbol $\alpha$ is a material parameter that amounts to the rate at which the mobile moisture becomes immobile, and $\beta$ represents the rate at which the immobile moisture becomes mobile. The Langmuir-type diffusion is discussed in detail in Appendix \ref{langmuir}
\section{Governing partial differential equations}\label{govequ}
The governing equations of the coupled material model expressed in the deformed consist of the balance of momentum to express mechanical behaviour and the Langmuir-type diffusion model to express moisture diffusion 
\begin{equation}
	\begin{aligned}
		{\rm{div}}\,\mathbf{T}\left(J,\bar{\mathbf{B}},\bar{\mathbf{B}}_e^j,m\right) = \mathbf{0} \hspace{2 mm} &\forall \hspace{2 mm} \mathbf{x}\in{\rm \Omega},\,\, {\rm and}\\[3mm]
		\dot{m}= D\,\Delta m_f = D\,{\rm div}\left({\rm grad}\left(m-m_b\right)\right)  \hspace{2 mm} &\forall \hspace{2 mm} \mathbf{x}\in{\rm \Omega},
	\end{aligned}
	\label{eq:numcop1}
\end{equation}
where the moisture-dependent Cauchy stress tensor $\mathbf{T}$ is expressed as
\begin{equation}
	\mathbf{T}\,=\,\mathbf{T}_{\rm vol}(J,\lambda_m,m) + \mathbf{T}_{\rm eq}({\mathbf{\bar{B}}},\lambda_m,m) + \sum_{j=1}^n\mathbf{T}_{\rm neq}^j({\mathbf{B}}_e^j,\lambda_m,m).
\end{equation}
The Cauchy stress is calculated with the local moisture concentration-dependent stiffness parameters to evaluate the ageing behaviour. The moisture-dependent stiffness parameters are calculated by interpolating the dry and saturated states of the material \cite{Sharma}. The interpolation is given by
\begin{equation}
	\mu(m)\,=\,f(m)\mu^{\rm{dry}}+(1-f(m))\mu^{\rm{sat}},
	\label{eq:numc}
\end{equation}
where $\mu(m)$ is the local stiffness parameters calculated at the integration points. $\mu^{\rm{dry}}$ and $\mu^{\rm{sat}}$ are the stiffness parameters of dry and saturated material samples. $f(m)$ is an interpolation function to couple mechanical and diffusion equations. The coupling function is defined using an exponential decay 
\begin{equation}
	f(m)\,=\,{\rm{exp}}(-\Lambda m),
	\label{eq:numcop}
\end{equation}
where $\Lambda$ is a parameter of the decay function. The coupling function has to satisfy the condition $0\le f(m) \le 1$, where $f(m)=1$ defines a dry state and $f(m)\approx 0$ defines a saturated state. 
\subsection{Boundary and initial conditions}
The initial and boundary conditions are needed for the defined governing equations of the coupled system of equations. The initial conditions are defined over the material domain to solve the Langmuir-type diffusion model. Total moisture and immobile moisture distribution in the material at time $t_0 = 0$ is applied as the initial boundary condition
\begin{equation}
	m\left(\mathbf{x},t=0\right) = 0,\hspace{2mm} m_b \left(\mathbf{x},t=0\right) = 0.
\end{equation}
Let Dirichlet and Neumann boundaries for the moisture diffusion are $\partial{\rm \Omega}_D$ and $\partial{\rm \Omega}_N$ and the mechanical problem are $\partial{\rm\Omega}_{\mathbf{u}}^D$ and $\partial{\rm \Omega}_{\mathbf{u}}^N$ and has to satisfy
\begin{equation}
	\begin{aligned}
		&\partial{\rm \Omega}_N\cup\partial{\rm \Omega}_D\,=\,\partial {\rm \Omega}, \,\hspace{3mm} \partial{\rm \Omega}_N\cap\partial{\rm \Omega}_D\,= \varnothing \\
		&\partial{\rm\Omega}_{\mathbf{u}}^D\cup\partial{\rm \Omega}_{\mathbf{u}}^N\,=\,\partial {\rm \Omega}, \,\hspace{3mm} \partial{\rm\Omega}_{\mathbf{u}}^D\cap\partial{\rm \Omega}_{\mathbf{u}}^N\,= \varnothing 
	\end{aligned}
	\label{conditions}
\end{equation}
The diffusion and deformation boundary conditions are defined over the specified Dirichlet and Neumann boundaries as follows
\begin{equation}
	\begin{aligned}
		\mathbf{u}\left(\mathbf{x},t\right) = \mathbf{u}_D\left(\mathbf{x},t\right) \,\, {\rm on} \,\,\partial{\rm \Omega}_{\mathbf{u}}^D \,\, &{\rm and} \,\, \mathbf{T}\cdot\mathbf{n} = \mathbf{t} \,\, {\rm on} \,\,\partial{\rm \Omega}_{\mathbf{t}}^N,\\
		m\left(m(\mathbf{x},t)\right) = m^{\rm eq}\hspace{1mm} \forall \hspace{1mm}\mathbf{x} \in \partial{\rm \Omega}_{m}^D \,\, &{\rm and} \,\, \mathbf{q}\left(\mathbf{x},t\right)= D\,{\rm grad}m_f\cdot \mathbf{n} \hspace{1mm}\forall\hspace{1mm}\mathbf{x} \in \partial{\rm \Omega}_{\mathbf{q}}^N.
	\end{aligned}
\end{equation}
where $\mathbf{t}$ is the traction on the surface $\partial{\rm \Omega}_{\mathbf{t}}^N$ with the normal vector $\mathbf{n}$, $m^{\rm eq}$ is the relative humidity in the surrounding atmosphere and $n$ is the outward normal vector on the boundary. The coupled problem is implemented and solved in deal.II finite element library \cite{DanielA,Bangerth,saad}. The displacement and diffusion fields of the coupled problem are solved individually as a coupled staggered field to obtain a stable implicit formulation.
\section{Numerical implementation and investigation}
The weak forms of the governing equations are derived to solve the partial differential equations defined in section \ref{govequ} using the finite element method. To this end, the arbitrary test functions $\delta \mathbf{u}$ are $\delta m$ are multiplied by the governing equations and are integrated over the material volume
\begin{equation}
	\begin{aligned}
		\int\limits_{\rm \Omega}\delta\mathbf{u}\cdot{\rm div}\,\mathbf{T}\left(\bar{\mathbf{B}},\bar{\mathbf{B}}_e^j,J,m\right){\rm dV} \, &= \,\mathbf{0},\\
		\int\limits_{\rm \Omega}\delta m\, \dot{m} {\rm dV} - \int\limits_{\rm \Omega}\delta m\,D\, {\rm div}\left({\rm grad}\left(m-m_b\right)\right){\rm dV} &= 0.
	\end{aligned}
	\label{impulrefstat}
\end{equation}
Finally, integration by parts leads to
\begin{equation}
	\begin{aligned}
		\mathbf{{r}}_{\mathbf{u}} (\mathbf{u}) = 
		\int\limits_{\rm \Omega} \mathbf{T}\left(\bar{\mathbf{B}},\bar{\mathbf{B}}_e^j,J,m\right)\colon{\rm grad}^s\,\delta\mathbf{u} \,{\rm dV} - \int\limits_{\partial\rm \Omega}\mathbf{T}\cdot\delta\mathbf{u}\, {\rm dA} &=\mathbf{0},\\
		{{\rm r}}_{m} ({m}) = \int\limits_{\rm \Omega}\delta m\, \dot{m}\, {\rm dV} +\int\limits_{\rm \Omega}\left[{\rm grad}\delta m\cdot D\,{\rm grad}\left(m-m_b\right)\right]{\rm dV} &= 0.
	\end{aligned}
\end{equation}
The diffusion is a long time process for a material to reach the equilibrium state. Therefore, the diffusion equation needs to consider a larger time step to solve the problem with less computational effort. A second-order time derivative proposed by Crank-Nicolson \cite{crank} is used to solve the diffusion problem since both the explicit and implicit time derivatives have temporal truncation errors for larger time steps \cite{MAZUMDER2016219}. The time discretisation with Crank-Nicolson method leads to the residual ${{\rm r}}_{m} ({m})$ 
\begin{equation}
	\begin{aligned}
		\!\!\!\!{{\rm r}}_{m} ({m}) = \int\limits_{\rm \Omega}\!\delta m\, \frac{m^{t+1}-m^t}{\Delta t} {\rm dV} &+ \int\limits_{\rm \Omega}\!\left[{\rm grad}\delta m\cdot D\,\frac{1}{2} {\rm grad}\left(m^{t+1}-m_b^{t+1}\right)\right]{\rm dV}\\
		&-\int\limits_{\rm \Omega}\!\left[{\rm grad}\delta m\cdot D\,\frac{1}{2} {\rm grad}\left(m^{t}-m_b^{t}\right)\right]{\rm dV} = 0,
	\end{aligned}
	\label{weakla4-1}
\end{equation}
where $\left(\bullet\right)^{t+1}$ and $\left(\bullet\right)^{t}$ are the values of the field variables calculated at current time $t+1\, \rm s$ and previous time $t\,\rm s $ steps. The evolution equation \eqref{eq:diff2} is solved to evaluate the immobile moisture concentration as
\begin{equation}
	\frac{m_b^{t+1}-m_b^t}{\Delta t} = \frac{1}{2}\left[\alpha \left(m^{t+1} - m^t\right)\right] - \frac{1}{2} \left[\left(\alpha + \beta\right)\left(m_b^{t+1}- m_b^t \right)\right]. 
\end{equation}
The differential equation is treated to consider the geometrical non-linearity because of the large deformations. The linearised approximation of the non-linear governing equations is solved with Newton's method using 
\begin{align}
	\mathbf{R}\left(\boldsymbol{\rm \Xi} + \Delta \boldsymbol{\rm \Xi}\right) \approx
	\mathbf{R}(\boldsymbol{\rm \Xi}) + {\rm D}_{ \Delta \boldsymbol{\rm \Xi}} \mathbf{R}\left(\boldsymbol{\rm \Xi}\right)\cdot{\rm d}\boldsymbol{\rm \Xi}=0,
	\label{Newton}
\end{align}
where ${\rm D}_{ \Delta \boldsymbol{\rm\Xi}} \left(\bullet \right)$ represents the directional derivative, also known as the spatial tangent tensor, that describes the change in the residuals $\mathbf{R}(\boldsymbol{\rm \Xi})$ in the direction of the unknown vector $\boldsymbol{\rm \Xi}$. The component of the directional derivative $\mathbf{{K}}^{m m}$ known as the diffusive matrix is 
\begin{equation}
	\mathbf{{K}}^{m m} = \int\limits_{\rm \Omega}{\rm grad}\,\delta m\, {\rm grad}\,\delta m\, {\rm dV},
\end{equation}
and the direction derivative component $\mathbf{{K}}^{ \mathbf{uu}}$ in the direction $\Delta \mathbf{u}$ is 
\begin{equation}
	\begin{split}
		\mathbf{K}^{\mathbf{uu}}={\rm D}_{\Delta \mathbf{u}} \mathbf{{r}}(\mathbf{u})
		&=
		\int\limits_{\rm \Omega}
		{\rm D}_{\Delta \mathbf{u}} \left(\mathbf{T}\left(\bar{\mathbf{B}},\bar{\mathbf{B}}_e^j,J,m\right)\right)  :
		{\rm grad}^s \, \delta \mathbf{u}
		\, {\rm dV}
		\\
		& +
		\int\limits_{\rm \Omega}
		\mathbf{T}\left(\bar{\mathbf{B}},\bar{\mathbf{B}}_e^j,J,m\right) :
		\left[
		{\rm Grad} \, \delta \mathbf{u} \cdot
		{\rm D}_{\Delta \mathbf{u}} \mathbf{F}^{-1}
		\right] {\rm dV},
	\end{split}
	\label{eq:tangent_pre}
\end{equation}
the directional derivative $\mathbf{{K}}^{ \mathbf{uu}}$ is simplified to 
\begin{equation}
	\begin{aligned}
		\mathbf{K}^{\mathbf{uu}}={\rm D}_{\Delta \mathbf{u}} \mathbf{{r}}(\mathbf{u})
		&=
		\int\limits_{\rm \Omega} {\rm grad}^s\, \Delta \mathbf{u} :\, \overset{4}{\boldsymbol{\kappa}}\left(\bar{\mathbf{B}},\bar{\mathbf{B}}_e^j,J,m\right)\, : {\rm grad}^s \, \delta \mathbf{u} \, {\rm dV} \\
		&+
		\int\limits_{\rm \Omega} {\rm grad}\,\delta \mathbf{u} : \left[ {\rm grad}\, \Delta \mathbf{u} \cdot \mathbf{T}\left(\bar{\mathbf{B}},\bar{\mathbf{B}}_e^j,J,m\right)
		\right]
		{\rm dV}.
	\end{aligned}
	\label{eqtangent}
\end{equation}
Here, the tangent $\overset{4}{\boldsymbol{\kappa}}$ is calculated as the sum of the volumetric $\overset{4}{\boldsymbol{\kappa}}\!{}_{\rm vol}$ and isochoric components composing of the equilibrium $\overset{4}{\boldsymbol{\kappa}}{}_{\rm eq}\left(\bar{\mathbf{B}},m\right)$ and $j=1,2,...,n$ non-equilibrium $\overset{4}{\boldsymbol{\kappa}}_{\rm neq}\left(\bar{\mathbf{B}}_e^j,m\right)$ parts of the viscoelastic model. The equation \eqref{tangent} in the Appendix defines the tangent matrices. $\bar{\mathbf{B}}_e^j$ of the $j^{th}$ non-equilibrium part is calculated from the evolution equation of the inelastic right Cauchy-Green deformation tensor $\bar{\mathbf{C}}_i^j$. This evolution equation is solved with the implicit Euler method in time combined with the local Newton method in space at every integration point.

The parameters of the mechanical and diffusion models are identified by fitting simulation curves to the experimental curves. A gradient-free method proposed by Nelder \& Mead \cite{Nelder} is used for parameter identification. At first, the Langmuir-type diffusion parameters are identified from the gravimetric tests conducted with 98\% r.H. in the atmosphere at an isothermal condition of $60\, ^\circ\rm C$. Then, the viscoelastic behaviour is studied with the uniaxial tensile tests performed on the dry and aged samples at an isothermal condition. The corresponding material parameters of the micromechanical model are identified for dry and aged samples. The micromechanical model parameters of dry and aged samples are evaluated to define the decay function to interpolate the material parameters to analyse the ageing behaviour.
\subsection{Diffusion}
A thin sample of $0.833\,{\rm mm}$ thick sample is used to investigate the moisture diffusion in the polyurethane adhesive. The top and bottom faces along thickness are exposed to moisture, whereas the other faces are isolated to surroundings so that the moisture diffusion is one dimensional. The Langmuir-type diffusion model parameters are identified with the curve fitting process. The optimum material parameters are listed in Table \ref{table:0}.  
\begin{table}[H]
	\caption{Langmuir-type diffusion parameters obtained from curve fitting method}
	\renewcommand{\arraystretch}{1.4}
	\begin{tabular}{l|l}
		\hline
		\multicolumn{2}{l}{Diffusion parameters at $60\,^\circ\,\rm C$}\\
		\hline
		diffusion coefficient $D\hspace{52 mm}$ & $ 7.925{\rm{E}}-05\, {\rm{ mm^2s^{-1}}}$\\
		rate at which immobile moisture becomes mobile again $\alpha$ & $2.727{\rm{E}}-05\,{\rm{s^{-1}}}$\\ 
		rate at which mobile moisture becomes immobile $\beta$ &  $2.247{\rm{E}}-03\,{\rm{s^{-1}}}$ \\
		\hline
	\end{tabular}
	\label{table:0}
\end{table}
The experimental and simulation results are plotted together to check for the deviation between the results. 
\begin{figure}[!htp]
	\centering
	\scalebox{0.8}{\input{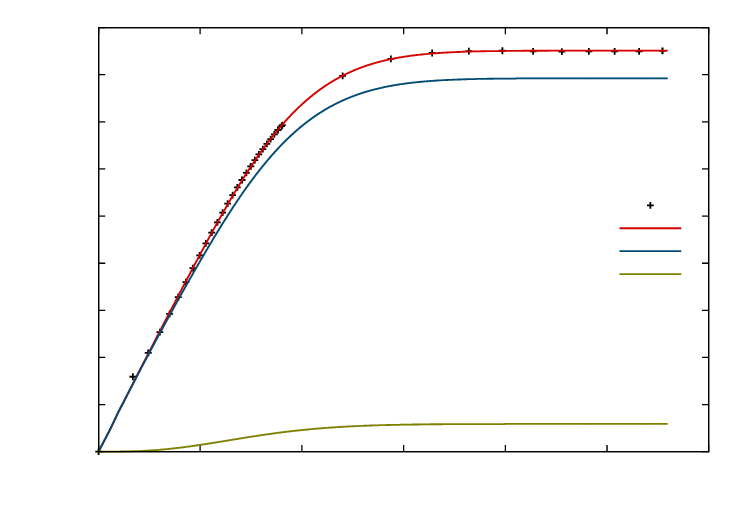}}
	\caption{Comparison between experimental (Exp) and simulation data of moisture absorption}
	\label{fig:difu}
\end{figure}

\begin{figure}[H]
	\centering
	\def\svgwidth{0.6\textwidth}
	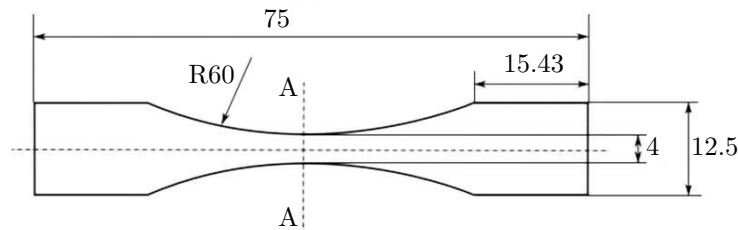
	\caption{Tailored tensile test samples with necked cross-section: all dimensions are in millimetres}
	\label{FIG:3}
\end{figure}

The total moisture concentration calculated with the material properties given in Table \ref{table:0} shows good agreement with the experimental results. The comparison between the experimental and simulation results is shown in Fig. \ref{FIG:3}.
\subsection{Material parameters of micromechanical model}
The viscoelastic behaviour of the crosslinked adhesive under the influence of moisture is investigated by performing a uniaxial tensile test. The test sample is optimised to have minimum cross-section at the centre. The motivation to use a tailored sample is to measure local strains at $2\,\rm mm$ span from the centre of the specimen. These tailored samples are not subjected to any pre-stressed or strains in the manufacturing process.

The samples are aged at different humid ($0\%$ r.H., $29\%$ r.H., $67\%$ r.H., $100\%$ r.H.) condition at an isothermal condition of $60\, ^\circ\rm C$ to investigate the moisture influence on the tensile strength. Seven Maxwell elements are used to model the viscoelastic behaviour. 
\begin{table}[H]
	\caption{Material parameters of micromechanical polymer network model at different ambient condition}
	\small\addtolength{\tabcolsep}{-3pt}
	\renewcommand{\arraystretch}{1.5}
	\begin{tabular}{|c|c|c|c|c|c|c|} 
		\hline
		\multicolumn{7}{|c|}{Material parameters of micromechanical model}                                                                                                                                                                                                                                     \\ \hline
		&                                                                  & \begin{tabular}[c]{@{}c@{}}Relaxation\\ times {[}s{]}\end{tabular} & $0\%$ r.H. & $29\%$ r.H. & $67\%$ r.H. & $100\%$ r.H. \\ \hline
		Equilibrium                                                                    & $c_{10}$ {[}MPa{]}                                               &                                                                    & 9.183      & 7.744       & 6.455       & 6.052        \\ \hline
		\multirow{7}{*}{Non-equilibrium}                                                & $c_{101}$ {[}MPa{]}                                              & 0.5                                                                & 5.223      & 4.654       & 4.225       & 4.044        \\ \cline{2-7} 
		& $c_{102}$ {[}MPa{]}                                              & 10                                                                 & 4.152      & 3.654       & 3.225       & 3.044        \\ \cline{2-7} 
		& $c_{103}$ {[}MPa{]}                                              & 100                                                                & 3.140      & 2.654       & 2.225       & 2.144        \\ \cline{2-7} 
		& $c_{104}$ {[}MPa{]}                                              & 500                                                                & 2.328      & 1.543       & 1.035       & 1.012        \\ \cline{2-7} 
		& $c_{105}$ {[}MPa{]}                                              & 1000                                                               & 1.582      & 1.317       & 0.618       & 0.404        \\ \cline{2-7} 
		& $c_{106}$ {[}MPa{]}                                              & 2500                                                               & 1.131      & 1.068       & 0.326       & 0.246        \\ \cline{2-7} 
		& $c_{107}$ {[}MPa{]}                                              & 5000                                                               & 0.961      & 0.778       & 0.686       & 0.107        \\ \hline
		\multirow{2}{*}{\begin{tabular}[c]{@{}c@{}}Wesslau \\ parameters\end{tabular}} & \begin{tabular}[c]{@{}c@{}}Average chain \\ stretch $^M\lambda_m$\end{tabular} &                                                                    & 1.194      & 1.287       & 1.658       & 1.931        \\ \cline{2-7} 
		& \begin{tabular}[c]{@{}c@{}}Polydispersity\\ index $Q$\end{tabular}   &                                                                    & 1.001      & 1.039       & 1.272       & 1.367        \\ \hline
	\end{tabular}
	\label{table1}
\end{table}
The finite element model is applied with the tensile boundary conditions with the micromechanical material parameters listed in Table \ref{table1} for different atmospheric conditions.

\begin{figure}[H]
	\centering
	\begin{tabular}{c c}		
		\hspace{-5mm}\subf{\scalebox{.65}{\input{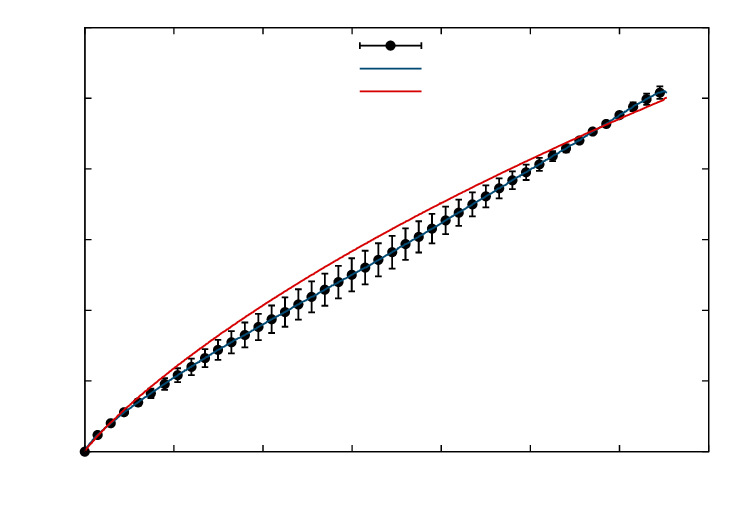}}}
		{\hspace{3mm}dry sample ($0\%$ r.H. ambient condition)}
		&
		\hspace{-5mm}\subf{\scalebox{.65}{\input{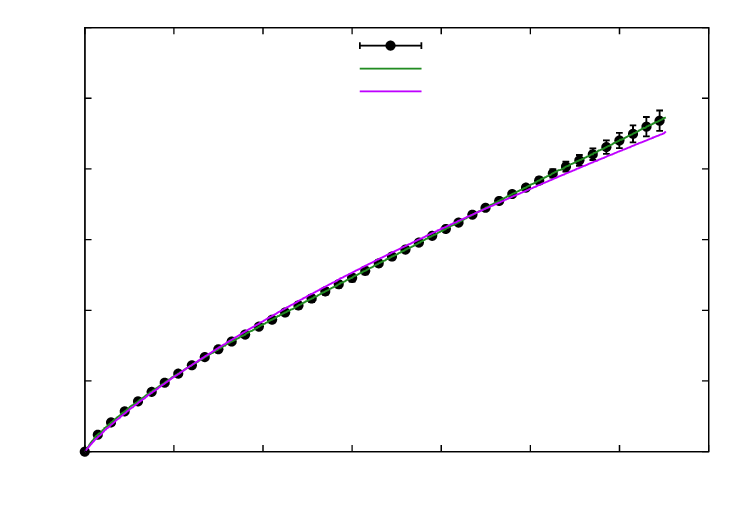}}}
		{\hspace{3mm}$29\%$ r.H. equilibrium moisture state}
		\\
		\\
		\hspace{-5mm}\subf{\scalebox{.65}{\input{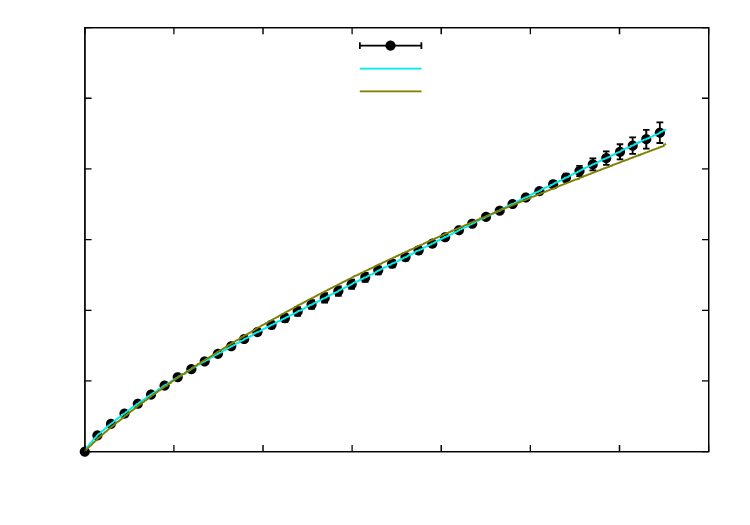}}}
		{\hspace{3mm}$67\%$ r.H. equilibrium moisture state}
		&
		\hspace{-5mm}\subf{\scalebox{.65}{\input{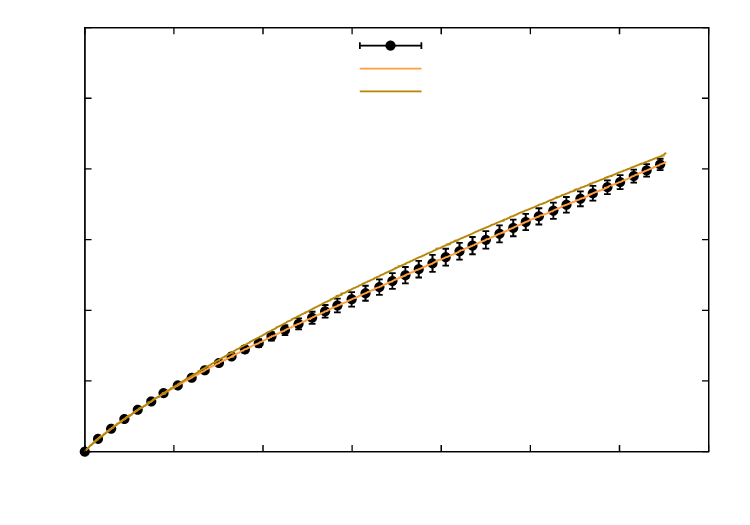}}}
		{\hspace{3mm}$100\%$ r.H. equilibrium moisture state}
	\end{tabular}
	\caption{Curve fitting of simulation data with experimental data under tensile test at different ambient moisture conditions at $60\,^\circ$C}
	\label{fig:abs}
\end{figure}

The finite element analysis performed with the optimal parameters is compared with the test results. Fig. \ref{fig:abs} shows the comparison of stress-stretch data between simulation and test results with the standard deviation. The tension test data plotted in the comparison corresponds to the mean values calculated from the test series consisting of five samples for aged samples at individual humid climatic condition. The standard deviation in the form of the error bar indicates that the problem is well-posed. 

It is apparent from the uniaxial tensile test that the tensile strength decreases with an increase in moisture concentration due to the decrease in the material stiffness. The stiffness parameters decrease exponentially with an increase in the local moisture concentration. Decay in the stiffness parameters is estimated by interpolating parameters as the function of moisture concentration \cite{Sharma}. The moisture-dependent stiffness parameters are calculated at integration points with equation \ref{eq:numc}. Dry $\mu^{\rm{dry}}$ and saturated $\mu^{\rm{sat}}$ at 100\% r.H. stiffness parameters listed in Table \ref{table1} are used in the interpolation. The coupling parameter $\Lambda$ of the integration function $f(m)$ given in equation \ref{eq:numcop} is identified as $\Lambda = 2.16$. 
\begin{figure}[!htp]
	\centering
	\scalebox{.9}{\input{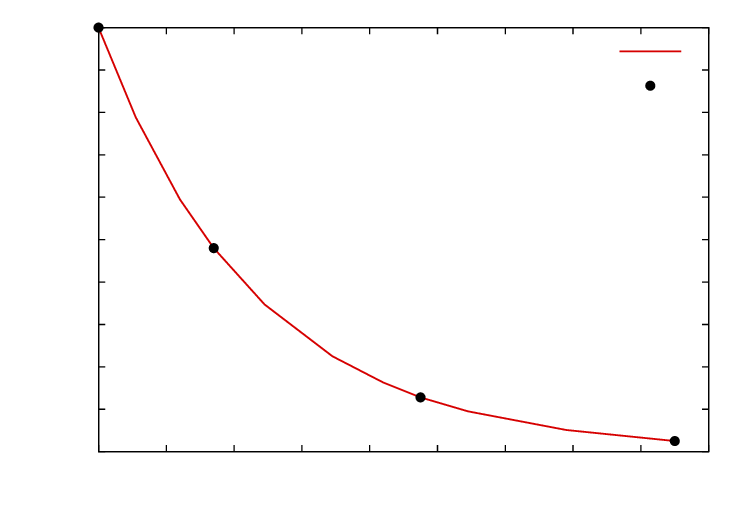}}
	\caption{Exponential decay function to interpolate the material parameters}
	\label{interpo} 
\end{figure}
\section{Investigation of coupled problem}
\begin{figure}[H]
	\centering
	\scalebox{.55}{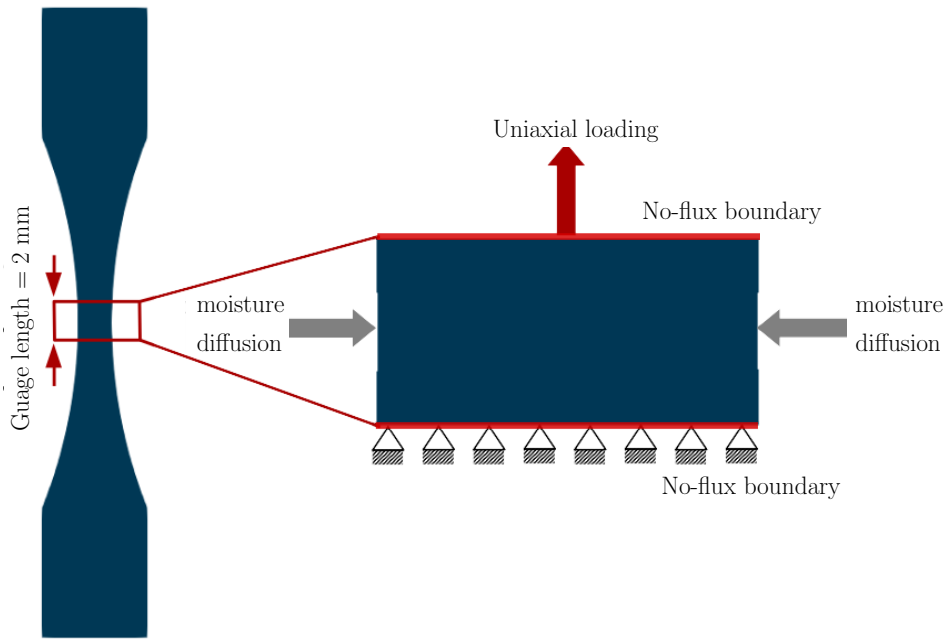}
	\caption{Left and right faces of $2\,\rm mm$ wide cross-section of the tailored tensile sample is subject to moisture diffusion for time $t\,\rm s$ with no-flux boundaries and then the sample is uniaxially loaded with a strain rate of $0.0005\,\rm s^{-1}$}
	\label{mechanicalbcs}
\end{figure}

Multi-physically coupled diffusion and deformation model is quantified by investigating the tailored sample with inhomogeneous moisture distribution. The sample is investigated for different ageing times exposed to $100\%$ r.H. atmospheric conditions. The aged samples for different times are applied with the tensile boundary conditions inconsistent with the uni-axial tensile tests. The stress-stretch data is measured locally at a span of $2\, \rm mm$ from the centre of the sample. For finite element analysis, the moisture is diffused from two end faces of the sample for time $t=4000 \rm s,\, 10000s,\, 15000s\, {\rm and} \,60000s$ to prepare aged samples with inhomogeneous moisture distribution. The aged samples is applied with the tensile test boundary condition. Fig. \ref{mechanicalbcs} shows the schematic representation of the applied boundary conditions. 

\begin{figure}[!htp]
	\centering
	\scalebox{.85}{\input{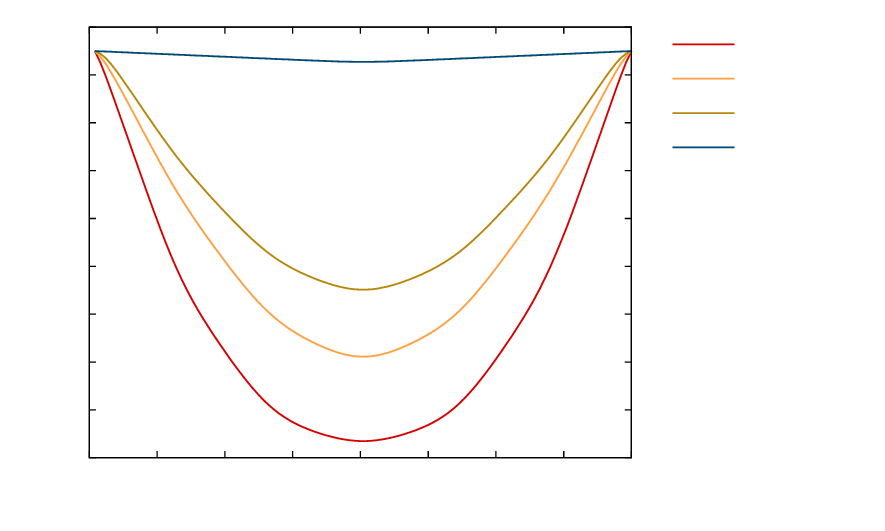}}
	\caption{Distribution of moisture over the cross-section of the adhesive sample determined by simulation at different times}
	\label{moisconc}
\end{figure}

Fig. \ref{moisconc} shows a comparison of moisture distribution along the centre of the sample at different times. Results indicate that the concentration of moisture molecules diffused in the sample increases with time and finally reaches to an equilibrium state.

\begin{figure}[H]
	\centering
	\scalebox{.78}{\input{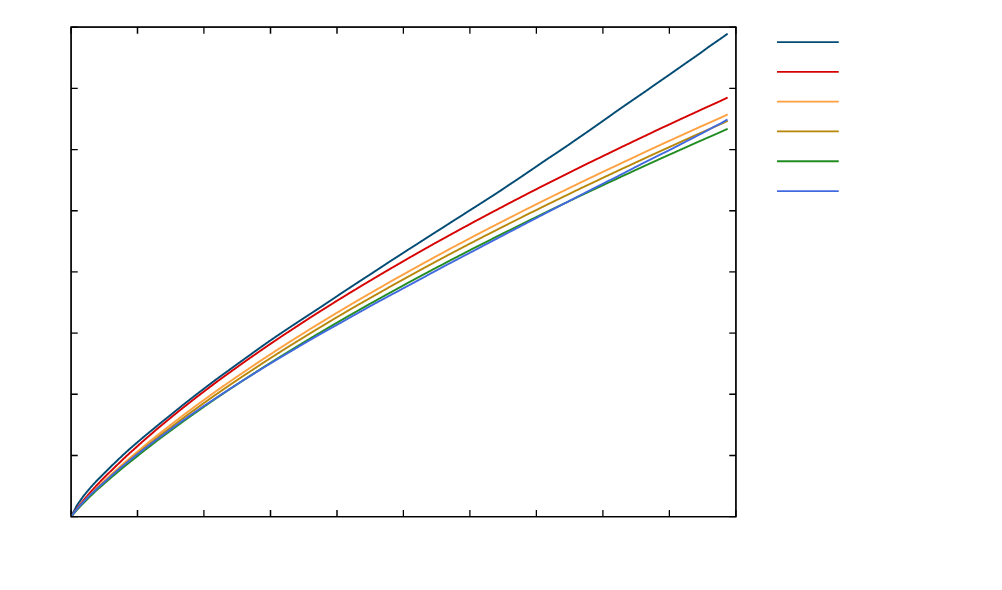}}
	\caption{Comparison of the stress-stretch curves of the samples with inhomogeneous moisture distribution with dry and saturated samples at $100\%$ relative humidity in climate}
	\label{coupdiff}
\end{figure}

The stiffness parameters at the integration points are calculated with dry and saturated material parameters (see table \ref{table1}) using the equation \eqref{eq:numc}. The experimental stretch-stress curves of dry and at $100\%$ r.H. saturated samples is plotted against the simulation results. It is evident from Fig. \ref{coupdiff} that the loss in the stiffness parameters with increasing moisture concentration results in lower stresses. The sample reaches the equilibrium at time $t = \rm \,60,000\,s$, as a result the simulation results coincides with the test results of the sample aged at 100\% r.H. The simulation results of sample with inhomogeneous moisture distribution falls between the the dry and 100\% r.H. aged samples validating the proposed coupled material model.
\section{Conclusion}
Experimental investigations performed on the crosslinked polyurethane adhesives show the ageing behaviour under moisture influence, which was not studied in detail. The key motivation of this work is due to the absence of the multi-physical coupled model between diffusion and viscoelastic behaviour. Thus, a micromechanical model is developed to analyse the finite-strain viscoelastic behaviour by considering the network evolution concept in modelling. The network evolution theory is used, so that the softening behaviour is modelled without defining damage function. The moisture diffusion shows anomalous behaviour in this crosslinked polyurethane adhesive, thus characterising the diffused moisture into mobile and immobile moisture concentration. Langmuir-type diffusion model is used in this work to model moisture diffusion. The micromechanical model is coupled with Langmuir-type diffusion by considering moisture-dependent parameters for the mechanical model. The moisture-dependent material parameters are calculated by interpolating the dry and saturated sample parameters. An exponential function is used in the interpolation of materials  at the integration points. The stress-stretch data calculated from the numerical analysis with coupled material model is compared with the experimental results. Even though this model displays an agreeable fit, the swelling behaviour in modelling is not accounted for. Therefore, it is necessary to incorporate deformation due to swelling in future work.  

\section*{Acknowledgments}
The research project 19730 N "Berechnung des instation\"aren mechanischen Verhaltens von alternden Klebverbindungen unter Einfluss von Wasser auf den Klebstoff" of the research association Forschungsvereinigung Stahlanwendung e.V. (FOSTA), D\"usseldorf was supported by the Federal Ministry of Economic Affairs and Energy through the AiF as part of the program for promoting industrial cooperative research (IGF) on the basis of a decision by the German Bundestag. The experimental investigations are carried out by the project associates from Lehrstuthl f\"ur Angewandte Mechanik, Universit\"at des Saarlandes and Fraunhofer-Institut f\"ur Fertigungstechnik und Angewandte Materialforschung.

\section*{Conflicts of interest}
The authors declare no conflicts of interest.

	

\newpage
\appendix
\section{Decomposition of Micromechanical energy}
The softening-based micromechanical model is an extension of the model to model viscoelastic behaviour is discussed in section \ref{micromech} and \ref{viscoelas}. The decomposition of the viscoelastic free energy into shape and volume-changing parts are discussed in detail to evaluate the constitutive equations.
\subsection{Isochoric part of free energy}
The invariants of the deformation tensor are necessary for modelling the micromechanical model. As aforementioned, invariants of the isochoric and volumetric parts discussed before being substituted in the equation \eqref{energy} to formulate the isochoric part of free energy
\begin{equation}
	\begin{aligned}
		W_{\rm{iso}}(\bar{\rm{I}}_1,\lambda_m) &= {\rm\Psi}_{\rm{iso}}(\bar{\rm{I}}_1) \left(\int\limits_{1}^{\infty}g_{\rm{iso}}(\lambda_m){\rm{d}} \lambda_m -\int\limits_{1}^{\lambda_c} g_{\rm{iso}}(\lambda_m){\rm{d}} \lambda_m \right),
		\label{eq:chain-1}
	\end{aligned}
\end{equation}
where the isotropic free energy of material ${\rm\Psi}_{\rm{iso}}$ is defined with Neo-Hookean model for simplicity \cite{Rivlin1948}. As a result, the isochoric part of the micromechanical free energy is
\begin{equation}
	\begin{aligned}
		W_{\rm{iso}}(\bar{\rm{I}}_1,\lambda_m) = c_{10} \left( \bar{{\rm{I}}}_1-3 \right) \left(1 -\int\limits_{1}^{\lambda_c} g_{\rm{iso}}(\lambda_m){\rm{d}} \lambda_m \right) ={\rm\Psi}_{\rm{iso}}(\bar{\rm{I}}_1) \left(1 -G_{\rm{iso}}\left(\lambda_c(\bar{\rm{I}}_1)\right) \right).
		\label{eq:chain-2}
	\end{aligned}
\end{equation}
Due to the implicit dependence of $G_{\rm{iso}}\left(\lambda_c(\bar{\rm{I}}_1)\right)$ on $\lambda_c(\bar{\rm{I}}_1)$, the necessary derivatives are computed using the chain rule to derive stress and tangent tensors. With the derivative operators $\frac{\partial(\cdots)}{\partial\lambda_c}=\dot{(\cdots)}$ and $\frac{\partial(\cdots)}{\partial \rm I_1}=(\cdots)'$, the first and second order derivatives are
\begin{equation}
	\begin{aligned}
		G'_{\rm{iso}}\left(\lambda_c(\bar{\rm{I}}_1)\right) = \frac{\partial G_{\rm{iso}}}{\partial\lambda_c}\frac{\partial\lambda_c}{\partial \bar{\rm{I}}_1} &=\dot{G}_{\rm{iso}}\lambda_c',\\[3mm]
		G''_{\rm{iso}}\left(\lambda_c(\bar{{\rm{I}}}_1)\right) = \frac{\partial}{\partial \bar{{\rm{I}}}_1}\left(\dot{G}_{\rm{iso}}\lambda_c'\right) &= \frac{\partial \dot{G}_{\rm{iso}}}{\partial\lambda_c}\frac{\partial\lambda_c}{\partial \bar{{\rm{I}}}_1}\lambda_c'+ \dot{G}_{\rm{iso}}\frac{\partial\lambda_c'}{\partial\bar{{\rm{I}}}_1}\\[+3mm]
		&= \ddot{G}_{\rm{iso}}\lambda_c'^2+\dot{G}_{\rm{iso}}\lambda_c''.
	\end{aligned}
	\label{eq:chain-3}
\end{equation}
The derivative of isochoric cumulative density distribution ${G}_{\rm{iso}}$ with respect to actual chain stretch $\lambda_c$ follows:
\begin{equation}
	\dot{G}_{\rm{iso}}\left(\lambda_c(\bar{{\rm{I}}}_1)\right) = \frac{a_0\, a_2}{\left(\lambda_c-1\right)}\rm{exp}\left(\frac{1}{4\, a_1} - \frac{\left(1+2\, a_1 \rm{ln}\left(a_2 (\lambda_c-1) \right) \right)^2}{4\, a_1}\right),
	\label{eq:chain-5}
\end{equation}
and the second order derivative of isochoric cumulative density distribution ${G}_{\rm{iso}}$ with respect to actual chain stretch $\lambda_c$ is calculated as 
\begin{equation}
	\begin{aligned}
		\ddot{G}_{\rm{iso}}\left(\lambda_c(\bar{\rm{I}}_1)\right) = \frac{a_0\, a_2}{\left(\lambda_c-1\right)^2}&\rm{exp}\left(\frac{1}{4\, a_1} - \frac{\left(1+2\, a_1 \rm{ln}\left(a_2 (\lambda_c-1) \right) \right)^2}{4\, a_1}\right) \left(2+2\, a_1 \rm{ln}\left(a_2 (\lambda_c-1) \right) \right).
	\end{aligned}
	\label{eq:chain-6}
\end{equation}
The first and second order derivative of current chain stretch $\lambda_c\left(\bar{\rm{I}}_1 \right)$ with respect to the first invariant of the isochoric Cauchy-Green deformation tensor $\bar{\rm{I}}_1$ are 
\begin{equation}
	\begin{aligned}
		\lambda_c'\left(\bar{\rm{I}}_1 \right) &= \frac{1}{6}\left( \frac{\bar{\rm{I}}_1}{3}\right)^{-1/2} = \frac{1}{6}\lambda_c^{-1/2},\hspace{5mm}
		\lambda_c''\left(\bar{\rm{I}}_1 \right) &= -\frac{1}{36}\left( \frac{\bar{\rm{I}}_1}{3}\right)^{-3/2} = \frac{1}{36}\lambda_c^{-3/2}
	\end{aligned}
	\label{eq:chain-8}
\end{equation}
The computation of the parameters $a_0$, $a_1$ and $a_2$ are given in the equation \eqref{param}. By combining equation \eqref{eq:chain-5}, \eqref{eq:chain-6} and \eqref{eq:chain-8}, the first and second order derivatives of free energy $W_{\rm{iso}}$ are evaluated to derive the stress and tangent tensors. The respective derivatives of the free energy function are
\begin{equation}
	\begin{aligned}
		W_{\rm{iso}}' &= G_{\rm{iso}}{\rm\Psi}_{\rm{iso}}' + G_{\rm{iso}}'{\rm\Psi}_{\rm{iso}} = G_{\rm{iso}} {\rm\Psi}_{\rm{iso}}' + \dot{G}_{\rm{iso}}\lambda_c'{\rm\Psi}_{\rm{iso}},\\[+3mm]
		W_{\rm{iso}}'' &= G_{\rm{iso}}{\rm\Psi}_{\rm{iso}}'' +2G_{\rm{iso}}'{\rm\Psi}_{\rm{iso}}' + G_{\rm{iso}}''{\rm\Psi}_{\rm{iso}}\\[+3mm]
		&= G_{\rm{iso}} {\rm\Psi}_{\rm{iso}}'' + 2\dot{G}_{\rm{iso}}\lambda_c'{\rm\Psi}_{\rm{iso}} + \left(\ddot{G}_{\rm{iso}}\lambda_c'^2 +\dot{G}_{\rm{iso}}\lambda_c'' \right){\rm\Psi}_{\rm{iso}}.
	\end{aligned}
\end{equation}
\subsection{Volumetric  part of free energy}
Due to the nearly incompressible behaviour of the material, the volumetric part of free energy density is dependent on  the volume ratio. With these considerations, the free energy function is based on the Jacobian $J$ and the actual chain stretch
\begin{equation}
	\begin{aligned}
		W_{\rm{vol}}(J,\lambda_m) &= {\rm\Psi}_{\rm{vol}}(J)\left(\int\limits_{1}^{\infty}g_{\rm{vol}}(\lambda_m)\rm{d} \lambda_m -\int\limits_{1}^{\lambda_c}g_{\rm{vol}}(\lambda_m)\rm{d} \lambda_m \right),
		\label{eq:chain-10}
	\end{aligned}
\end{equation}
where the volumetric part of free energy is a simple quadratic function of Jacobian $J$ 
\begin{equation}
	{\rm\Psi}_{\rm{vol}}(J) = \frac{1}{D}(J-1)^2,\hspace{1mm} {\rm with}\hspace{1mm} J=\sqrt{\bar{{\rm{I}}}_3},
	\label{volfree}
\end{equation}
where $D$ is a material constant. By substituting equation \eqref{volfree} in \eqref{eq:chain-10} leads to the volumetric part of the micromechanical free energy
\begin{equation}
	\begin{aligned}
		W_{\rm{vol}}(J,\lambda_m) &= \frac{1}{D}(J-1)^2\left(	1-\int\limits_{1}^{\lambda_c}g_{\rm{vol}}(\lambda_m)\rm{d} \lambda_m\right)\\[+3mm]
		&= \frac{1}{D}(J-1)^2(1-G_{\rm{vol}}\left(\lambda_c(J) \right).
	\end{aligned}
\end{equation}
First-order derivative of volumetric cumulative density distribution function $G'_{\rm{vol}}\left(\lambda_c(J)\right)$ is derived because of the implicit dependence on the Jacobian $J$
\begin{equation}
	G'_{\rm{vol}}\left(\lambda_c(J)\right) = \frac{\partial G_{\rm{vol}}}{\partial\lambda_c}\frac{\partial\lambda_c}{\partial J} =\dot{G}_{\rm{vol}}\lambda_c' \hspace{1mm} {\rm where} \hspace{1mm} \lambda_c(J) = J^{\sfrac{1}{3}}.
	\label{eq:chain-12}
\end{equation}
The second order derivative of the volumetric form of cumulative density distribution function $G''_{\rm{vol}}\left(\lambda_c(J)\right)$ can be derived by applying the chain rule
\begin{equation}
	\begin{aligned}
		G''_{\rm{vol}}\left(\lambda_c(J)\right) = \frac{\partial}{\partial J}\left(\dot{G}_{\rm{vol}}\lambda_c'\right) &= \frac{\partial \dot{G}_{\rm{vol}}}{\partial\lambda_c}\frac{\partial\lambda_c}{\partial J}\lambda_c'+ \dot{G}_{\rm{vol}}\frac{\partial\lambda_c'}{\partial J}\\[+3mm]
		&= \ddot{G}_{\rm{vol}}\lambda_c'^2+\dot{G}_{\rm{vol}}\lambda_c''.
		\label{eq:chain-13}
	\end{aligned}
\end{equation}
$\dot{G}_{\rm{vol}}\left(\lambda_c(J)\right)$ given in the equation \eqref{eq:chain-12} is computed as the partial derivative of ${G}_{\rm{vol}}\left(\lambda_c(J)\right)$ with current chain stretch $\lambda_c(J)$
\begin{equation}
	\dot{G}_{\rm{vol}}\left(\lambda_c(J)\right) = \frac{a_0 a_2}{\left(\lambda_c-1\right)}\rm{exp}\left(\frac{1}{4\, a_1} - \frac{\left(1+2 a_1 \rm{ln}\left(a_2 (\lambda_c-1) \right) \right)^2}{4\, a_1}\right),
	\label{eq:chain-14}
\end{equation}
and $\ddot{G}_{\rm{vol}}\left(\lambda_c(J)\right)$ required in derivation of equation \eqref{eq:chain-13} is
\begin{equation}
	\begin{aligned}
		\ddot{G}_{\rm{vol}}\left(\lambda_c(J)\right) = \frac{a_0 a_2}{\left(\lambda_c-1\right)^2}&\rm{exp}\left(\frac{1}{4\, a_1} - \frac{\left(1+2 a_1 \rm{ln}\left(a_2 (\lambda_c-1) \right) \right)^2}{4\, a_1}\right) \left(2+2 a_1 \rm{ln}\left(a_2 (\lambda_c-1) \right) \right).
	\end{aligned}
	\label{eq:chain-15}
\end{equation}
First and second order derivatives of the current chain stretch $\lambda_c\left(J \right)$ needed to derive equation \eqref{eq:chain-12} and \eqref{eq:chain-13} are
\begin{equation}
	\begin{aligned}
		\lambda_c'\left(J \right) &= \frac{1}{3}{J}^{-1/3} = \frac{1}{3}\lambda_{\rm{vol}}^{-1},\hspace{5mm}
		\lambda_c''\left(J \right) &= -\frac{1}{9}{J}^{-4/3} = -\frac{1}{9}\lambda_{\rm{vol}}^{-4}.
	\end{aligned}
	\label{eq:chain-17}
\end{equation}
To evaluate of stress and tangent tensors of the desired material model, one must derive first and second-order derivatives of the free energy function. The first- and second order derivatives of the volumetric part of micromechanical free energy density $W_{\rm{vol}}$ are
\begin{equation}
	\begin{aligned}
		W_{\rm{vol}}' &= G_{\rm{vol}}{\rm\Psi}_{\rm{vol}}' + G_{\rm{vol}}'{\rm\Psi}_{\rm{vol}} = G_{\rm{vol}} {\rm\Psi}_{\rm{vol}}' + \dot{G}_{\rm{vol}}\lambda_c'{\rm\Psi}_{\rm{vol}}\\[+3mm]
		W_{\rm{vol}}'' &= G_{\rm{vol}}{\rm\Psi}_{\rm{vol}}'' +2G_{\rm{vol}}'{\rm\Psi}_{\rm{vol}}' + G_{\rm{vol}}''{\rm\Psi}_{\rm{vol}}\\
		&= G_{\rm{vol}} {\rm\Psi}_{\rm{vol}}'' + 2\dot{G}_{\rm{vol}}\lambda_c'{\rm\Psi}_{\rm{vol}} + \left(\ddot{G}_{\rm{vol}}\lambda_c'^2 +\dot{G}_{\rm{vol}}\lambda_c'' \right){\rm\Psi}_{\rm{vol}}.
	\end{aligned}
	\label{eq:chain-16-2}
\end{equation}
\section{Stress and Elasticity tensors}\label{stress}
The stress tensor in the updated Lagrangian formulation is obtained by applying a push-forward operation over the second Piola-Kirchhoff stress tensor. The stress tensor involved in the updated Lagrangian formulation is the Kirchhoff stress tensor $\boldsymbol{\tau}$ or the Cauchy stress tensor $\mathbf{T}$. In here, the Cauchy stress tensor $\mathbf{T}$ is calculated from the Kirchhoff stress tensor $\boldsymbol{\tau}$, where the Kirchhoff stresses $\boldsymbol{\tau}$ are 
\begin{equation}
	\begin{aligned}
		\!\!\!\!\boldsymbol{\tau}_{\rm eq}\left({\bar{\mathbf{B}}},\lambda_{m}\right) &= 2\mathbf{B}\cdot\frac{\partial W_{\rm eq} \left(\bar{\rm I}_1^{\bar{\mathbf{B}}},\lambda_{m} \right)}{\partial\mathbf{B}} = 2J^{-2/3}\left(W_{\rm eq}\right)'\left(\bar{\mathbf{B}} - \frac{1}{3}\bar{\rm I}_1^{\bar{\mathbf{B}}}\mathbf{I}\right);\\[3mm]
		\!\!\!\!\boldsymbol{\tau}_{\rm vol}\left(J,\lambda_{m}\right) &= 2\mathbf{B}\cdot\frac{\partial W_{\rm vol} \left(J,\lambda_{m} \right)}{\partial\mathbf{B}} = JW_{\rm vol}\mathbf{I};\\[3mm]
		\!\!\!\!\boldsymbol{\tau}^{j}_{\rm neq}\!\!\left({\bar{\mathbf{B}}^j_e},\lambda_{m}\right) &= 2\bar{\mathbf{B}}^j_e\cdot\frac{\partial W_{\rm neq}^j \left(\bar{\rm I}_1^{\bar{\mathbf{B}}^j_e},\lambda_{m} \right)}{\partial\bar{\mathbf{B}}^j_e} = 2J^{-2/3}\!\left(W_{\rm neq}^j\right)'\!\!\left(\bar{\mathbf{B}}^j_e - \frac{1}{3}\bar{\rm I}_1^{\bar{\mathbf{B}}^j_e}\mathbf{I}\right)\!\!.
		\label{stressten}
	\end{aligned}
\end{equation}
and the Cauchy stresses are calculated 
\begin{equation}
	\mathbf{T} = J^{-1}\boldsymbol{\tau}.
\end{equation}
In non-linear finite element simulation the solution is often solved incrementally using Newton's method and computation of the tangent is crucial for the solution. The tangent tensor of the softening micromechanical free energy is
\begin{equation}
	\begin{aligned}
		\overset{4}{\boldsymbol{\kappa}} &= 4\mathbf{B}\cdot\frac{\partial^2 W \left(\bar{\rm I}_1^{\bar{\mathbf{B}}}, \bar{\rm I}_1^{\bar{\mathbf{B}}^j_e},J,m \right)}{\partial\mathbf{B}\,\partial\mathbf{B}}\cdot\mathbf{B} =\,\,\, \overset{4}{\boldsymbol{\kappa}}_{\rm eq} + \overset{4}{\boldsymbol{\kappa}}_{\rm vol} + \overset{4}{\boldsymbol{\kappa}}_{\rm neq},\\[3mm]
		\overset{4}{\boldsymbol{\kappa}}_{\rm eq} = 4\mathbf{B}\cdot&\frac{\partial^2 W_{\rm eq} \left(\bar{\rm I}_1^{\bar{\mathbf{B}}}, m \right)}{\partial\mathbf{B}\, \partial\mathbf{B}}\cdot\mathbf{B}; \hspace{5mm} \overset{4}{\boldsymbol{\kappa}}_{\rm vol} =4\mathbf{B}\cdot\frac{\partial^2 W_{\rm vol} \left(J,m \right)}{\partial\mathbf{B}\,\partial\mathbf{B}}\cdot\mathbf{B},\\[3mm]
		\overset{4}{\boldsymbol{\kappa}}_{\rm neq} = 4\bar{\mathbf{B}}_e^j\cdot&\frac{\partial^2 W_{\rm neq}^j \left(\bar{\rm I}_1^{\bar{\mathbf{B}}^j_e},m \right)}{\partial\bar{\mathbf{B}}_e^j\, \partial\bar{\mathbf{B}}_e^j}\cdot\bar{\mathbf{B}}_e^j.
	\end{aligned}
\end{equation}
The individual components of the spatial elasticity tensor take the form
\begin{equation}
	\begin{aligned}
		\overset{4}{\boldsymbol{\kappa}}_{\rm eq} =\,& 4 \left(W_{\rm eq} \left(\bar{\rm I}_1^{\bar{\mathbf{B}}},m\right)\right)' \left( \mathbf{I}\otimes\bar{\mathbf{B}}\right)^{s_{24}} + 4 \left(W_{\rm eq} \left(\bar{\rm I}_1^{\bar{\mathbf{B}}},m\right)\right)'' \left(\bar{\mathbf{B}}\otimes\bar{\mathbf{B}}\right) - \\
		&\frac{4}{3}\left(\bar{\rm I}_1^{\bar{\mathbf{B}}} \left(W_{\rm eq} \left(\bar{\rm I}_1^{\bar{\mathbf{B}}},m\right)\right)''\!\! + \left(W_{\rm eq} \left(\bar{\rm I}_1^{\bar{\mathbf{B}}},m \right) \right)' \right) \left(\bar{\mathbf{B}}\otimes\mathbf{I}+\mathbf{I}\otimes\bar{\mathbf{B}}\right)+\\
		&\frac{4}{9}\left(\left(\bar{\rm I}_1^{\bar{\mathbf{B}}}\right)^2 \left(W_{\rm eq} \left(\bar{\rm I}_1^{\bar{\mathbf{B}}},m\right)\right)''\!\! + \bar{\rm I}_1^{\bar{\mathbf{B}}} \left(W_{\rm eq} \left(\bar{\rm I}_1^{\bar{\mathbf{B}}},m\right)\right)' \right)\left(\mathbf{I}\otimes\mathbf{I}\right)\\[3mm]
		\overset{4}{\boldsymbol{\kappa}}_{\rm vol} =\,& \left(J^2 \left(W_{\rm vol} \left(J,m \right)\right)'' + J \left(W_{\rm vol} \left(J,m \right)\right)' \right)\mathbf{I}\otimes\mathbf{I}\\[3mm]
		\overset{4}{\boldsymbol{\kappa}}_{\rm neq} =\,& 4 W'_{\rm neq} \left(\bar{\rm I}_1^{\bar{\mathbf{B}}^j_e}, m \right) \left(\mathbf{I} \otimes \bar{\mathbf{B}}\right)^{s_{24}} + 4 W''_{\rm neq} \left(\bar{\rm I}_1^{\bar{\mathbf{B}}^j_e}, m \right) \left( \bar{\mathbf{B}} \otimes\bar{\mathbf{B}}\right) - \\
		&\frac{4}{3}\left(\bar{\rm I}_1^{\bar{\mathbf{B}}^j_e} W''_{\rm neq} \left(\bar{\rm I}_1^{\bar{\mathbf{B}}^j_e}, m \right) + W'_{\rm neq} \left(\bar{\rm I}_1^{\bar{\mathbf{B}}^j_e}, m \right) \right)\left(\bar{\mathbf{B}}\otimes\mathbf{I}+\mathbf{I}\otimes\bar{\mathbf{B}}\right)+\\
		&\frac{4}{9}\left(\left(\bar{\rm I}_1^{\bar{\mathbf{B}}^j_e}\right)^2 W''_{\rm neq} \left(\bar{\rm I}_1^{\bar{\mathbf{B}}^j_e},m \right) + \bar{\rm I}_1^{\bar{\mathbf{B}}^j_e} W'_{\rm neq} \left(\bar{\rm I}_1^{\bar{\mathbf{B}}^j_e},m \right)\right)\left(\mathbf{I}\otimes\mathbf{I}\right).
	\end{aligned}
	\label{tangent}
\end{equation}
In the definition of the isochoric component of the equilibrium $W_{\rm eq}$ and non-equilibrium $W^j_{\rm neq}$ of $j^{th}$ Maxwell element free energy functions are motivated by the Neo-Hook energy function for its simplicity which is required to implement the viscoelastic material model based on micromechanical material model. Based on the fundamental idea discussed in the development of the viscoelastic material model, the micromechanical free energy functions take the form
\begin{equation}
	\begin{aligned}
		\!\!\!&{\rm \Psi}_{\rm eq}\left(\bar{{\rm I}}_1^{\bar{\mathbf{B}}}\right)=c_{10}\left(\bar{{\rm I}}_1^{\bar{\mathbf{B}}}-3\right) {\rm results \, in}\, W_{\rm eq} = {\rm \Psi}_{\rm eq}\left(\bar{{\rm I}}_1^{\bar{\mathbf{B}}}\right)G\left(\lambda_c\left(\bar{{\rm I}}_1^{\bar{\mathbf{B}}}\right)\right),\\[5mm]
		\!\!\!&{\rm \Psi}^j_{\rm neq}\left(\bar{{\rm I}}_1^{\bar{\mathbf{B}}_e^j}\right)=c_{10j}\left(\bar{{\rm I}}_1^{\bar{\mathbf{B}}_e^j}-3\right) {\rm results \, in}\, W^j_{\rm neq} = {\rm \Psi}_{\rm eq}\left(\bar{{\rm I}}_1^{\bar{\mathbf{B}}_e^j}\right)G\left(\lambda_c\left(\bar{{\rm I}}_1^{\bar{\mathbf{B}}_e^j}\right)\right),
	\end{aligned}
\end{equation}
and the free energy for the volumetric component of the equilibrium part is introduced with a quadratic function of Jacobian as follows
\begin{equation}
	\Psi_{\rm vol}\left(J\right)= \frac{1}{D}\left(J - 1\right)^2 {\rm results \, in}\, W_{\rm vol} = \Psi_{\rm vol} \left(J \right) G \left( \lambda_c \left(J \right) \right),
\end{equation}
where $D$ corresponds to the shear modulus. When $J=1$ the material corresponds to the incompressible material. The first and second-order derivatives of the volumetric free energy are given in the equation \eqref{eq:chain-16-2}
\section{Modelling transport of moisture}\label{langmuir}
The diffusion of moisture in the crosslinked polyurethane adhesive is modelled here with the Langmuir-type diffusion model \cite{CAR78}, so that the dispersion of moisture concentration into mobile and immobile moisture concentrations are taken into consideration. The Langmuir-type diffusion model is formulated as follows:
\begin{equation}
	\dot{m}= D\,\Delta m_f = D\,{\rm div}\left({\rm grad}\left(m-m_b\right)\right),
	\label{eq:diff1}
\end{equation}
where $D$ is diffusion coefficient, $m_f$ is the mobile moisture concentration, $m_b$ is immobile moisture concentration and $m = m_f+m_b$ is total moisture concentration. The moisture distribution along the material is characterised by $m=0$ for the dry state and $m=m^{\rm eq}$ for the equilibrium state. The diffusion equation \eqref{eq:diff1} is supplemented by an evolution equation to compute immobile moisture concentration $m_b$ 
\begin{equation}
	\dot{m_b}= \alpha m_f - \beta m_b.
\end{equation}
The simple idea behind the evolution equation is that moisture is bound faster when there is a large amount of mobile moisture. In contrast, when the bound moisture concentration is large, a significant amount of mobile moisture is released to become mobile. The two constants $\alpha$ and $\beta$ describe the time constants of these two effects. An equilibrium moisture distribution occurs when the time derivative of the immobile moisture concentration of the evolution equation becomes zero. As a result, the evolution equation leads to
\begin{equation}
	\alpha m_f^{\rm eq} = \beta m_b^{\rm eq}.
	\label{equilibrium}
\end{equation}
The mobile and immobile concentration at the equilibrium condition is evaluated by substituting total moisture concentration in equation \eqref{equilibrium}. The resulting mobile and immobile moisture concentrations are
\begin{equation}
	m_f^{\rm eq} = \frac{m^{\rm eq}}{1+\sfrac{\alpha}{\beta}},\hspace{1cm} m_b^{\rm eq} = \frac{m^{\rm eq} \sfrac{\alpha}{\beta}}{1+\sfrac{\alpha}{\beta}}.
\end{equation}
Both boundary and initial conditions are needed to solve the Langmuir model. The initial conditions are defined for the entire material volume $\rm \Omega$ at the time $t_0 = 0$ given
\begin{equation}
	m\left(\mathbf{x},t=0\right) = m\left(\mathbf{x}\right),\hspace{2mm} m_b \left(\mathbf{x},t=0\right) = m_{b_0}\left(\mathbf{x}\right),
\end{equation}
where the spatial functions of the total moisture concentration $m\left(\mathbf{x}\right)$ and the bound moisture concentration $m_{b_0}\left(\mathbf{x}\right)$ are used to define the initial condition of moisture distribution. The boundary conditions are discretised into Dirichlet and Neumann conditions. On the Dirichlet boundaries, $\partial{\rm \Omega}_{m}^D$ is defined with the concentration values as the Dirichlet boundary condition. In contrast, the Neumann boundary $\partial{\rm \Omega}_{\mathbf{q}}^N$ is applied with the moisture flux. The boundary surfaces need to satisfy the condition
\begin{equation}
	\partial{\rm \Omega}_{m}^D\cup \partial{\rm \Omega}_{\mathbf{q}}^N = \partial{\rm \Omega},\hspace{10mm} \partial{\rm \Omega}_{m}^D \cap \partial{\rm \Omega}_{\mathbf{q}}^N =\varnothing
\end{equation}
over the entire material volume $\rm \Omega$. The Dirichlet boundaries are applied with the moisture concentration as follows
\begin{equation}
	m\left(m(\mathbf{x},t)\right) = m^{\rm eq}\hspace{1mm} \forall \hspace{1mm}\mathbf{x} \in \partial{\rm \Omega}_{m}^D
\end{equation}
and the moisture flux is applied on the Neumann boundaries, and the Neumann boundary condition is given as:
\begin{equation}
	\mathbf{q}\cdot\mathbf{n} = D{\rm grad}m_f\cdot \mathbf{n} = \mathbf{q}\left(\mathbf{x},t\right)\hspace{1mm}\forall\hspace{1mm}\mathbf{x} \in \partial{\rm \Omega}_{\mathbf{q}}^N.
\end{equation}
As a result, the moisture flux takes the form
\begin{equation}
	\mathbf{q} = D\frac{\partial m_f}{\partial\mathbf{x}} = D{\rm grad}\, m_f
\end{equation}
where $\mathbf{n}$ is the normal outward vector on the boundary.
 
\end{document}